\documentclass{IEEEtran}
\usepackage{amsmath,amssymb,amsfonts}
\usepackage{algorithmic}
\usepackage{algorithm}
\usepackage{array}
\usepackage[caption=false,font=normalsize,labelfont=sf,textfont=sf]{subfig}
\usepackage{textcomp}
\usepackage{stfloats}
\usepackage{url}
\usepackage{verbatim}
\usepackage{graphicx}
\usepackage{cite}
\usepackage{xfrac}
\usepackage[bb=boondox]{mathalfa}
\usepackage{nicematrix}
\usepackage{siunitx}
\usepackage[
colorlinks = true,
linkcolor = blue!80!black,
urlcolor  = blue!80!black,
citecolor = blue!80!black,
anchorcolor = blue!80!black]{hyperref}

\usepackage{esint} 

\newcommand{\secref}[1]{Section~\ref{#1}}
\newcommand{\figref}[1]{Fig.~\ref{#1}}

\newcommand{\volintp}[1][]{\int_{\Omega_{#1}}d\Omega'}
\newcommand{\surfint}[1][]{\int_{\Gamma_{#1}}d\Gamma}
\newcommand{\surfintp}[1][]{\int_{\Gamma_{#1}}d\Gamma'}

\newcommand{\pvsurfintp}[1][]{\fint_{\Gamma_{#1}}d\Gamma'}
\newcommand{\pvint}{\fint}

\newcommand{\eps}{\varepsilon}
\newcommand{\abs}[1]{\lvert#1\rvert}

\newcommand{\curl}[1][]{\nabla\times{#1}}

\newcommand{\divp}[1][]{\nabla'\cdot{#1}}
\newcommand{\nhat}[1][]{\hat{n}_{#1}}

\newcommand{\lapl}[1][]{\nabla^2{#1}}

\newcommand{\matr}[1]{\mathbf{#1}}

\newcommand{\E}[1][]{\vec{E}_{#1}}
\newcommand{\Er}[1][]{\E[#1]\,(\vec{r})}
\newcommand{\Erp}[1][]{\E[#1]\,(\vec{r}\,')}

\renewcommand{\H}[1][]{\vec{H}_{#1}}
\newcommand{\Hr}[1][]{\H[#1]\,(\vec{r})}
\newcommand{\Hrp}[1][]{\H[#1]\,(\vec{r}\,')}

\newcommand{\J}[1][]{\vec{J}_{#1}}
\newcommand{\Jr}[1][]{\J[#1]\,(\vec{r})}
\newcommand{\Jrp}[1][]{\J[#1]\,(\vec{r}\,')}

\newcommand{\rhor}[1][]{\rho_{#1}\,(\vec{r})}
\newcommand{\rhorp}[1][]{\rho_{#1}\,(\vec{r}\,')}

\newcommand{\Grrpki}[1][]{G(\vec{r}, \vec{r}\,', k_{#1})}

\newcommand{\hmat}{\ensuremath{\mathcal{H}}-matrix}
\newcommand{\hmats}{\ensuremath{\mathcal{H}}-matrices}
\newcommand{\hlu}{\ensuremath{\mathcal{H}}-LU}

\def\BibTeX{{\rm B\kern-.05em{\sc i\kern-.025em b}\kern-.08em
    T\kern-.1667em\lower.7ex\hbox{E}\kern-.125emX}}

\begin{document}

\title{A Hierarchically Compressed Immittance Operator Formulation\\ for Fast Statics-to-Full-Wave Electromagnetic Simulation\\ with the Boundary Element Method}

\author{Shashwat Sharma, \IEEEmembership{Member,~IEEE}%
\thanks{Manuscript received Month X, 2026; revised Month Y, 2026.}%
\thanks{This research was supported by the School of Engineering at the University of Birmingham through a new-faculty start-up grant. \emph{(Corresponding author: Shashwat Sharma.)}}%
\thanks{Shashwat Sharma is with the Department of Electronic, Electrical and Systems Engineering, in the School of Engineering, at the University of Birmingham, Edgbaston, B15 2TT, United Kingdom (e-mail: s.sharma.11[at]bham.ac.uk).}%
}

\markboth{IEEE Transactions on Antennas and Propagation,~Vol.~x, No.~y, Month~2026}%
{S. Sharma: An Immittance Operator Formulation for Broadband Electromagnetic Simulation}

\maketitle

\begin{abstract}
The boundary element method allows the electromagnetic simulation of three-dimensional problems with a two-dimensional discretization, using a surface integral representation of Maxwell's equations.
For applications involving broadband simulations or multiscale geometries, the well-known low-frequency breakdown of conventional integral equation formulations must be addressed, while ensuring that dielectric, semiconducting, and lossy metallic regions are modeled rigorously, including the skin effect.
In this work, a new boundary element formulation based on an immittance operator is proposed, where the underlying integral equations are inherently valid from statics to the full-wave regime.
To enable efficient matrix assembly and solution, a hierarchical matrix compression strategy is devised, leveraging fast direct factorization as an intermediate step, and fast matrix-vector products for the iterative solution of the final system of equations.
The proposed formulation is capable of electrostatic, magnetostatic, quasistatic, and full-wave simulations of dielectrics, semiconductors, and lossy metals.
Representative numerical examples drawn from a variety of applications demonstrate the accuracy and efficiency of the proposed method.
\end{abstract}


\section{Introduction}\label{sec_intro}

\IEEEPARstart{T}{he} boundary element method, based on surface integral equations, has been extensively investigated for electromagnetic simulation in a variety of applications, ranging from radiation and scattering problems, to board, package, and chip-level components in the semiconductor industry~\cite{BookChew2009}.
These methods allow modeling three-dimensional structures with a two-dimensional surface mesh, without the need to discretize the volume enclosed by object surfaces.

Conventional boundary element formulations tend to become numerically unstable at low frequencies, and invalid in the static limit~\cite{Chew2000,Chew2008}.
Even at moderately high frequencies, multiscale structures may include features that are electrically small, which can lead to a poorly conditioned discretized system matrix~\cite{DJiao2014,Andriulli2021}.
Yet, broadband electromagnetic simulation of multiscale structures is essential in some applications, including signal and power integrity analysis of high-speed interconnects, and the design of metamaterials and metasurfaces.
In the former, geometric sizes can range from microns to millimeters in the same structure, and their network characteristics must often be extracted from DC to tens or hundreds of gigahertz.
In the latter, unit cells are sub-wavelength in size, but the entire structure can be electrically large.
Similar challenges may arise in the modeling of transmon qubits for quantum computing, where millimeter-scale microwave circuits transition to nanometer-scale superconducting elements~\cite{Oliver2019,Roth2023}.
The multiscale and/or broadband nature of these simulations also requires conductor losses and the skin effect to be captured accurately at low and high frequencies, and simple metal models such as impedance boundary conditions may not suffice~\cite{BookYuferevIda2010}.
Therefore, using the boundary element method for these problems calls for techniques to remedy its low-frequency breakdown, while modeling dielectrics, semiconductors, and lossy metals rigorously.

Techniques to address the low-frequency breakdown problem include the use of loop-star and quasi-Helmholtz decompositions~\cite{Vecchi1999,Chew2000,JFLee2003,Andriulli2012,Andriulli2013,Andriulli2017,Andriulli2023}, advanced preconditioning strategies~\cite{Andriulli2008,Andriulli2011,Andriulli2021}, decoupled reformulations of the integral equations~\cite{Greengard2018}, incorporating additional charge-like unknown quantities~\cite{Oijala2006,Chew2009,Volkov2014}, and deriving integral equations in terms of potentials rather than fields~\cite{Chew2014,Greengard2016,Shanker2019,AWPLVPIE,Shanker2022,MTTVPIE}.
Many of these works consider perfectly conducting or dielectric objects, but not rigorously modeled lossy conductors.
Formulations for multi-region problems, with lossy conductor modeling, include those based on object-wise matrix operators that can be interpreted as global impedance or admittance boundary conditions~\cite{Chew2007,QianChew2009,DJiao2013,Utkarsh2016,AWPLSLIM,Zutter2025}, which require dedicated acceleration strategies for large objects~\cite{DJiao2013,TAPAIMin}.
Other methods have modeled lossy conductors with the help of basis functions defined on a barycentric refinement of the original mesh~\cite{Chew2017,JMMCTSLDM,MTTVPIE}, which tends to increase matrix assembly times, and may lead to an inaccurate representation of charge density~\cite{Oijala2013}.

Many of the referenced works alleviate low-frequency breakdown but are still invalid at exactly~\qty{0}{\hertz}.
Some of these methods, such as augmented formulations for penetrable materials~\cite{QianChew2009}, still break down or become inaccurate at frequencies or electrical sizes that may not be easy to predict beforehand, and depend on the type of structure and application, as was shown in~\cite{JMMCTFSIE}.
Furthermore, DC-to-full-wave modeling can be important in applications such as the signal integrity analysis of high-speed interconnects, where frequency-domain scattering parameters must be analyzed in the time domain to study the instantaneous impedance of interconnects, which requires~{\qty{0}{\hertz}--\num{100}$+$\,\unit{\giga\hertz}} frequency sweeps.
Therefore, it is both scientifically interesting and practically useful to seek formulations that are inherently valid and numerically stable from~\qty{0}{\hertz} to the full-wave regime, while modeling dielectric, semiconducting, and lossy metallic regions rigorously, thereby reflecting the full generality of Maxwell's equations.

In~\cite{JMMCTFSIE}, the F-TENH/NETH formulation was introduced, which is accurate and stable from exactly~\qty{0}{\hertz} to the full-wave regime, and does not require combining equations associated with different materials.
This avoids a possible loss of precision due to large material contrasts, unlike other combined-field formulations~\cite{BookGurel2014}, and also avoids discretizations that require a barycentric refinement of the mesh~\cite{Chew2017,JMMCTSLDM,MTTVPIE}.
However, the F-TENH/NETH system of equations is twice as large as the closely-related augmented formulations in~\cite{QianChew2009,AWPLSLIM}, and approximately~\num{1.7}$\times$ larger than conventional combined-field formulations for penetrable materials~\cite{Oijala2005}.
Furthermore, the F-TENH/NETH system relies heavily on a sparse preconditioner of the same size, whose factorization can be correspondingly more expensive compared to preconditioning the augmented formulations in~\cite{QianChew2009,AWPLSLIM} or the conventional combined-field formulations~\cite{Oijala2005}.

A new formulation is presented here, which aims to substantially improve upon the one in~\cite{JMMCTFSIE}.
The novelty of this work lies in its two main contributions: 
\begin{itemize}
    \item A new frequency-stable immittance operator formulation is introduced, which retains the broadband modeling advantages of the F-TENH/NETH system~\cite{JMMCTFSIE}, but with half the system size, and correspondingly improved performance. 
    The immittance operator uses object-wise matrix factorizations to eliminate unknowns as an intermediate step, and is valid from~\qty{0}{\hertz} to the full-wave regime, unlike related admittance and impedance operator approaches~\cite{Chew2007,Utkarsh2016,AWPLSLIM,Zutter2025}.
    \item A multi-region hierarchical ($\mathcal{H}$)-matrix~\cite{BookHackbusch2014} compression strategy is devised for the proposed formulation, to provide nearly log-linear time and memory complexity from~\qty{0}{\hertz} to the full-wave regime.
    The matrices associated with the immittance operator are cast in a form that allows leveraging a fast direct hierarchical LU~(\hlu{}) solver~\cite{BookHackbusch2014} for the intermediate object-wise factorizations, while the final system of equations is solved iteratively with fast matrix-vector products.
    This avoids the need for nested iterative solvers, unlike existing approaches~\cite{TAPAIMin}.
\end{itemize}
The proposed formulation is capable of fast electrostatic, magnetostatic, quasistatic, and full-wave simulations of dielectrics, semiconductors, and lossy metals, as shown in \secref{sec_res}.

The paper is organized as follows: \secref{sec_fsie} reviews the frequency-stable integral equations on which the proposed method is built.
\secref{sec_io} introduces the new immittance operator formulation, and \secref{sec_acc} describes the proposed~\hmat{} compression strategy.
Numerical results demonstrating accuracy and scaling are presented in \secref{sec_res}, and concluding remarks are provided in \secref{sec_concl}.

\section{Frequency-Stable Surface Integral Equations}\label{sec_fsie}

Time-harmonic fields are assumed with a time dependence of~$e^{j\omega t}$, where~$\omega$ is the angular frequency, and~${j=\sqrt{-1}}$.
The structure to be simulated is assumed to consist of~$N_{\mathrm{obj}}$ disjoint homogeneous objects, and the background is also assumed to be homogeneous.
Object~$i$ is denoted as~$\Omega_i$, and its bounding surface~$\Gamma_i$ is assumed to be closed, for~$i\in\{1,2,\ldots,N_{\mathrm{obj}}\}$.
The background region is indicated by~$\Omega_0$, and~$\Gamma_0=\bigcup_{i=1}^{N_{\mathrm{obj}}}\Gamma_i$.

The frequency-stable integral equations of~\cite{JMMCTFSIE} are used as a starting point, which are derived from the Helmholtz equations for the electric and magnetic fields,
\begin{align}
    \lapl{\Er} + k_i^2\,\Er &= j\omega\mu_i\,\Jr[\mathrm{im}]
        + \frac{1}{\eps_i}\nabla\rhor[\mathrm{im}],\label{Ehelm}\\
    \lapl{\Hr} + k_i^2\,\Hr &= -\curl{\Jr[\mathrm{im}]},\label{Hhelm}
\end{align}
where~$\vec{r}\in\Omega_i$, and~${i\in\{0,1,\ldots,N_{\mathrm{obj}}\}}$.
In~\eqref{Ehelm} and~\eqref{Hhelm},~$\eps_i$ and~$\mu_i$ are the permittivity and permeability associated with~$\Omega_i$, respectively, and the wave number~$k_i$ is defined via
\begin{align}
  k_i^2 = -(j\omega\mu_i)\,(j\omega\eps_i + \sigma_i),\label{kdef}
\end{align}
where~$\sigma_i$ is the electrical conductivity associated with~$\Omega_i$.
Additionally,~$\Jr[\mathrm{im}]$ and~$\rhor[\mathrm{im}]$ are impressed volume current and charge densities, respectively, and it is assumed that there are no sources on~$\Gamma_i$.

Applying the vector-dyadic Green's second theorem~\cite{BookHanson2002} gives the electric and magnetic field surface integral representations proposed by Stratton and Chu in~\cite{StrattonChu1939}, where, crucially, both tangential and normal components of the electric and magnetic field appear as surface-bound sources.
Letting~$\vec{r}\to\Gamma_i$ from within~$\Omega_i$, and taking the tangential trace, rotated tangential trace, and divergence of these surface integral representations provides the set of six frequency-stable surface integral equations used as the foundations of the proposed formulation and the one in~\cite{JMMCTFSIE},
\begin{multline}
    -\nhat[i]\times\left\{\left(\frac{1}{2}\,\mathcal{I} + \nhat[i]\times\left[\curl\mathcal{S}_{i}\right]_{\mathrm{pv}}\right)\left[\nhat[i]'\times\Erp\right]
    \right.
    \\
    \left.
    - j\omega\mu_i\,\nhat[i]\times\mathcal{S}_i\left[\nhat[i]'\times\Hrp\right]
    - \nhat[i]\times\nabla\mathcal{S}_i\left[\nhat[i]'\cdot\Erp\right]
    \right\}
    \\
    = 
    -\nhat[i]\times\nhat[i]\times\left\{-j\omega\mu_i\,\mathcal{V}_i\left[\Jrp[\mathrm{im}]\right](\vec{r})
    \right.
    \\
    \left.
    - \frac{1}{\eps_i}\,\nabla\mathcal{V}_i\left[\rhorp[\mathrm{im}]\right](\vec{r})\right\},\label{tefie}
\end{multline}
\begin{multline}
    \left(\frac{1}{2}\,\mathcal{I} + \nhat[i]\times\left[\curl\mathcal{S}_{i}\right]_{\mathrm{pv}}\right)\left[\nhat[i]'\times\Erp\right]
    \\
    - j\omega\mu_i\,\nhat[i]\times\mathcal{S}_i\left[\nhat[i]'\times\Hrp\right]
    - \nhat[i]\times\nabla\mathcal{S}_i\left[\nhat[i]'\cdot\Erp\right]
    \\
    = 
    \nhat[i]\times\left\{-j\omega\mu_i\,\mathcal{V}_i\left[\Jrp[\mathrm{im}]\right](\vec{r})
    \right.
    \\
    \left.
    - \frac{1}{\eps_i}\,\nabla\mathcal{V}_i\left[\rhorp[\mathrm{im}]\right](\vec{r})\right\},\label{nefie}
\end{multline}
\begin{multline}
    \mathcal{S}_i\left[\divp\left(\nhat[i]'\times\Hrp\right)\right]
    + (j\omega\eps_i + \sigma_i)\,\mathcal{S}_i\left[\nhat[i]'\cdot\Erp\right]
    = 0,\label{defie}
\end{multline}
\begin{multline}
    -\nhat[i]\times\left\{\left(\frac{1}{2}\,\mathcal{I} + \nhat[i]\times\left[\curl\mathcal{S}_{i}\right]_{\mathrm{pv}}\right)\left[\nhat[i]'\times\Hrp\right]
    \right.
    \\
    \left.
    + (j\omega\eps_i + \sigma_i)\,\nhat[i]\times\mathcal{S}_i\left[\nhat[i]'\times\Erp\right]
    \right.
    \\
    \left.
    - \nhat[i]\times\nabla\mathcal{S}_i\left[\nhat[i]'\cdot\Hrp\right]
    \right\}
    \\
    =
    -\nhat[i]\times\nhat[i]\times\curl\mathcal{V}_i\left[\Jrp[\mathrm{im}]\right](\vec{r}),\label{tmfie}
\end{multline}
\begin{multline}
    \left(\frac{1}{2}\,\mathcal{I} + \nhat[i]\times\left[\curl\mathcal{S}_{i}\right]_{\mathrm{pv}}\right)\left[\nhat[i]'\times\Hrp\right]
    \\
    + (j\omega\eps_i + \sigma_i)\,\nhat[i]\times\mathcal{S}_i\left[\nhat[i]'\times\Erp\right]
    - \nhat[i]\times\nabla\mathcal{S}_i\left[\nhat[i]'\cdot\Hrp\right]
    \\
    =
    \nhat[i]\times\curl\mathcal{V}_i\left[\Jrp[\mathrm{im}]\right](\vec{r}),\label{nmfie}
\end{multline}
\begin{multline}
    \mathcal{S}_i\left[\divp\left(\nhat[i]'\times\Erp\right)\right]
    - j\omega\mu_i\,\mathcal{S}_i\left[\nhat[i]'\cdot\Hrp\right]
    = 0,\label{dmfie}
\end{multline}
for~$\vec{r}\in\Gamma_i$,~$i\in\{0,1,\ldots,N_{\mathrm{obj}}\}$, where primed coordinates represent sources on~$\Gamma_i$, and~$\nhat[i]$ is the outward unit normal vector on~$\Gamma_i$.
The integral operators in~\eqref{tefie}--\eqref{dmfie} are
\begin{align}
    \mathcal{S}_i\left[\widetilde{x}(\vec{r}\,')\right](\vec{r}) &= \surfintp[i]\,\Grrpki[i]\,\widetilde{x}(\vec{r}\,'),\\
    \mathcal{V}_i\left[\widetilde{x}(\vec{r}\,')\right](\vec{r}) &= \volintp[i]\,\Grrpki[i]\,\widetilde{x}(\vec{r}\,'),
\end{align}
where~$\widetilde{x}(\vec{r}\,')$ may be a scalar or a vector, and the scalar Green's function kernel is
\begin{align}
    \Grrpki[i] = \frac{e^{-jk_i\,\abs{\vec{r} - \vec{r}\,'}}}{4\pi\,\abs{\vec{r} -  \vec{r}\,'}}.
    \label{hgf}
\end{align}
In~\eqref{tefie}--\eqref{dmfie},~$\mathcal{I}$ is the identity operator, and~$\left[\cdot\right]_\mathrm{pv}$ indicates an integration in the principal value sense.
Equations~\eqref{tefie}--\eqref{dmfie} are the frequency-stable integral equations~\cite{JMMCTFSIE}, which remain valid as~${\omega\to0}$, and in particular when~${\omega=0}$.

\section{Immittance Operator Formulation}\label{sec_io}

Using~\eqref{tefie}--\eqref{dmfie} as the foundations, the goal of this work is to derive a set of object-wise operators~$\mathcal{X}_i$ that relate~$
{\nhat[i]'\times\Hrp}$ and~${\nhat[i]'\cdot\Erp}$ to~$
{\nhat[i]'\times\Erp}$ and~${\nhat[i]'\cdot\Hrp}$ as
\begin{align}
    \begin{bmatrix}
        \nhat[i]'\times\Erp \\ \nhat[i]'\cdot\Hrp
    \end{bmatrix}
    =
    \underbrace{\begin{bmatrix}
        \mathcal{Z}_i & \mathcal{P}_i \\
        \mathcal{Q}_i & \mathcal{Y}_i
    \end{bmatrix}}_{\displaystyle\triangleq\mathcal{X}_i}
    \begin{bmatrix}
        \nhat[i]'\times\Hrp \\ \nhat[i]'\cdot\Erp
    \end{bmatrix}\label{xdef}
\end{align}
for~${i\in\{1,2,\ldots,N_{\mathrm{obj}}\}}$.
This allows eliminating~$
{\nhat[i]'\times\Erp}$ and~${\nhat[i]'\cdot\Hrp}$ as unknowns from the final system of equations, leading to a smaller system matrix compared to the~{F-TENH/NETH} formulation~\cite{JMMCTFSIE}.
In~\eqref{xdef},~$\mathcal{Z}_i$ resembles a global impedance operator, similar to the one in~\cite{Chew2007}, while~$\mathcal{Y}_i$ resembles a global admittance operator, similar to the concept introduced in~\cite{Zutter2005}, but for normal field components rather than tangential ones.
The off-diagonal operators~$\mathcal{P}_i$ and~$\mathcal{Q}_i$ relate normal and tangential field components to each other.
Due to the mixed impedance-admittance nature of~$\mathcal{X}_i$, it is referred to as an \emph{immittance} operator, following the terminology used in~\cite{BookBode1945,Matthaei1954} to refer to quantities that can denote both an impedance and an admittance.

A key distinction between the immittance operator sought here, and existing impedance and admittance operator formulations~\cite{Zutter2005,Chew2007,Utkarsh2016,AWPLSLIM}, is that the former is free of low-frequency breakdown and can be solved from exactly~$0\,$Hz to the full-wave regime, as confirmed in \secref{sec_res}.
This is a direct consequence of the use of the frequency-stable integral equations~\eqref{tefie}--\eqref{dmfie}, as described next in Sections~\ref{subsec_bc}--\ref{subsec_fin}.

\subsection{Interface Conditions and Discretization}\label{subsec_bc}

First, interface conditions are set on each~$\Gamma_i$ following the approach in~\cite{JMMCTFSIE}.
Tangential field continuity on~$\Gamma_i$ implies
\begin{align}
  \nhat[0]'\times\Erp\bigl|_{\Gamma_i^+} +\, \nhat[i]'\times\Erp\bigl|_{\Gamma_i^-} &= 0,\label{nxebc}\\
  \nhat[0]'\times\Hrp\bigl|_{\Gamma_i^+} +\,\nhat[i]'\times\Hrp\bigl|_{\Gamma_i^-} &= 0,\label{nxhbc}
\end{align}
for~$i\in\{1,2,\ldots,N_{\mathrm{obj}}\}$, where~$\Gamma_i^+$ and~$\Gamma_i^-$ represent~$\vec{r}\,'$ approaching~$\Gamma_i$ from~$\Omega_0$ or~$\Omega_i$, respectively.
For the normal component of the magnetic field, the relation used is
\begin{align}
  \mu_0\,\nhat[0]'\cdot\Hrp\bigl|_{\Gamma_i^+} +\,\mu_i\,\nhat[i]'\cdot\Hrp\bigl|_{\Gamma_i^-} = 0,\label{ndhbc}
\end{align}
for~$i\in\{1,2,\ldots,N_{\mathrm{obj}}\}$.
The interface relation for the normal component of the electric field is defined, for each~${i\in\{1,2,\ldots,N_{\mathrm{obj}}\}}$, via the following condition.
\begin{itemize}
\item If~$\sigma_0=0$ and~$\sigma_i=0$, the interface condition used is
  \begin{align}
    \eps_0\,\nhat[0]'\cdot\Erp\bigl|_{\Gamma_i^+} +\,\eps_i\,\nhat[i]'\cdot\Erp\bigl|_{\Gamma_i^-} = 0.\label{ndebc1}
  \end{align}
\item If~$\sigma_0>0$, and/or~$\sigma_i>0$, the interface condition used is
  \begin{multline}
    (j\omega\eps_0 + \sigma_0)\,\nhat[0]'\cdot\Erp\bigl|_{\Gamma_i^+} \\+ (j\omega\eps_i + \sigma_i)\,\nhat[i]'\cdot\Erp\bigl|_{\Gamma_i^-} = 0.\label{ndebc2}
  \end{multline}
\end{itemize}

Using~\eqref{nxebc}--\eqref{ndebc2}, the unknowns either on~$\Gamma_i^+$ or on~$\Gamma_i^-$ can be eliminated, for~${i\in\{1,2,\ldots,N_{\mathrm{obj}}\}}$.
In this work, for~\eqref{nxebc}--\eqref{ndhbc}, unknowns on~$\Gamma_i^-$ are eliminated.
However, for the normal component of the electric field, the choice is made independently for each~${i\in\{1,2,\ldots,N_{\mathrm{obj}}\}}$, depending on~$\sigma_0$ and~$\sigma_i$~\cite{JMMCTFSIE}:
\begin{itemize}
\item If~${\sigma_i\neq 0}$, eliminate~${\nhat[i]'\cdot\Erp\bigl|_{\Gamma_i^-}}$ using
  \begin{align}
    \nhat[i]'\cdot\Erp\bigl|_{\Gamma_i^-} = -\frac{(j\omega\eps_0 + \sigma_0)}{(j\omega\eps_i + \sigma_i)}\nhat[0]'\cdot\Erp\bigl|_{\Gamma_i^+}.\label{ndebc3}
  \end{align}
\item If~${\sigma_0 \neq 0}$ and~${\sigma_i = 0}$, eliminate~${\nhat[0]'\cdot\Erp\bigl|_{\Gamma_i^+}}$ using
  \begin{align}
    \nhat[0]'\cdot\Erp\bigl|_{\Gamma_i^+} = -\frac{(j\omega\eps_i)}{(j\omega\eps_0 + \sigma_0)}\nhat[i]'\cdot\Erp\bigl|_{\Gamma_i^-}.\label{ndebc4}
  \end{align}
\item If~${\sigma_0 = \sigma_i = 0}$, use~\eqref{ndebc1} to eliminate~${\nhat[i]'\cdot\Erp\bigl|_{\Gamma_i^-}}$ with
  \begin{align}
    \nhat[i]'\cdot\Erp\bigl|_{\Gamma_i^-} = -\frac{\eps_0}{\eps_i}\nhat[0]'\cdot\Erp\bigl|_{\Gamma_i^+}.\label{ndebc5}
  \end{align}
\end{itemize}

With these interface conditions, discretization proceeds by generating a planar triangular mesh for the structure.
The unknown surface-bound sources in~\eqref{tefie}--\eqref{dmfie} are expanded in terms of finite-support basis functions on the mesh.
The vector unknowns~$\nhat[i]'\times\Erp$ and~$\nhat[i]'\times\Hrp$ are expressed in terms of edge-length-normalized Rao-Wilton-Glisson~(RWG) functions~\cite{RWG1982} associated with mesh edges, while area-normalized piecewise constant functions associated with mesh triangles are used for the scalar unknowns~$\nhat[i]'\cdot\Erp$ and~$\nhat[i]'\cdot\Hrp$.
The divergence-conforming RWG functions ensure continuity across mesh edges of the surface current-like quantities~$\nhat[i]'\times\Erp$ and~$\nhat[i]'\times\Hrp$.
Column vectors~$\matr{H}_i^{(\mathrm{t})}$,~$\matr{E}_i^{(\mathrm{t})}$,~$\matr{E}_i^{(\mathrm{n})}$, and~$\matr{H}_i^{(\mathrm{n})}$ store the unknown expansion coefficients of~${\nhat[i]'\times\Hrp}$,~${\nhat[i]'\times\Erp}$,~${\nhat[i]'\cdot\Erp}$ and~${\nhat[i]'\cdot\Hrp}$, respectively, associated with~$\Omega_i$, for~${i\in\{1,2,\ldots,N_{\mathrm{obj}}\}}$.
The expansion coefficients in these column vectors can correspond to fields on~$\Gamma_i^+$ or~$\Gamma_i^-$ based on the elimination rules defined above and in~\eqref{ndebc3}--\eqref{ndebc5}.
The concatenation of expansion coefficients over all objects is denoted without a subscript, as~$\matr{H}^{(\mathrm{t})}$,~$\matr{E}^{(\mathrm{t})}$,~$\matr{E}^{(\mathrm{n})}$, and~$\matr{H}^{(\mathrm{n})}$, respectively.
Without loss of generality, the rules in~\eqref{ndebc3} and~\eqref{ndebc5} are assumed hereafter for simplicity, but the scenario in~\eqref{ndebc4} can easily be incorporated into the proposed formulation on a per-object basis.

\subsection{Discrete Immittance Operator}\label{subsec_int}

For the interior regions~$\Omega_{i}$,~${i\in\{1,2,\ldots,N_{\mathrm{obj}}\}}$, with the unknown expansions described in \secref{subsec_bc}, two discretized surface integral equations are chosen to model the interior physics: the first is a linear combination of~\eqref{nefie} and~\eqref{tmfie} tested with edge-length-normalized RWG functions, and the second is~\eqref{dmfie} tested with area-normalized triangle-wise constant functions.
Assuming there are no sources inside the object, a pair of discretized equations is obtained,
\begin{multline}
    \begin{bmatrix}
        \nu\,\widetilde{\matr{K}}_{i}^{(\mathrm{r})} + \widetilde{\eps}_i\,\matr{L}_i^{(\mathrm{t})}
        &
        b_i^{(\mathrm{h})}\,\matr{D}_i^{\mathrm{T}}\matr{L}_i^{(\mathrm{s})}\matr{B}_i^{(\mathrm{h})}
        \\
        \matr{F}_i^{(\mathrm{h})}\matr{L}_i^{(\mathrm{s})}\matr{D}_i
        & -\widetilde{\mu}_i\,b_i^{(\mathrm{h})}\,\matr{F}_i^{(\mathrm{h})}\matr{L}_i^{(\mathrm{s})}\matr{B}_i^{(\mathrm{h})}
    \end{bmatrix}
    \begin{bmatrix}
        \matr{E}_{i}^{(\mathrm{t})} \\
        \matr{H}_{i}^{(\mathrm{n})}
    \end{bmatrix}
    \\+
    \begin{bmatrix}
        \nu\,\widetilde{\mu}_i\,\matr{L}_i^{(\mathrm{r})} + \widetilde{\matr{K}}_{i}^{(\mathrm{t})}
        &
        \nu\,b_i^{(\mathrm{e})}\,\matr{M}_{i}
        \\
        \matr{0}
        &
        \matr{0}
    \end{bmatrix}
    \begin{bmatrix}
        \matr{H}_{i}^{(\mathrm{t})} \\
        \matr{E}_{i}^{(\mathrm{n})}
    \end{bmatrix}
    =
    \begin{bmatrix}
        \matr{0} \\ \matr{0}
    \end{bmatrix},\label{sysin}
\end{multline}
where the entries of the discretized integral operator matrices are defined in the Appendix.
In~\eqref{sysin}, the following symbols have been introduced for brevity:
\begin{align}
    \widetilde{\mu}_i \triangleq j\omega\mu_i,\\
    \widetilde{\eps}_i \triangleq j\omega\eps_i + \sigma_i.
\end{align}
The scalar~$\nu$ is a weighting parameter that balances accuracy with matrix conditioning, as discussed in~\cite{JMMCTFSIE}.
In this work,~${\nu=0.05}$ was used where~$i$ corresponds to a good conductor, and~${\nu=1}$ was used otherwise.
As in~\cite{JMMCTFSIE}, this choice of~$\nu$ was found to be robust and did not require tuning.
Interface conditions are taken into account in~\eqref{sysin} by defining~${b_i^{(\mathrm{h})}=\sfrac{\mu_0}{\mu_i}}$ using~\eqref{ndhbc},
and~$b_i^{(\mathrm{e})}$ using~\eqref{ndebc3}--\eqref{ndebc5}. 

The left-hand side of~\eqref{sysin} has been designed to resemble the magnetic-field dual of augmented electric field formulations~\cite{Chew2009}, which have been shown to avoid low-frequency breakdown provided a suitable constraint is placed on the charge-like scalar unknown~${\nhat[i]'\cdot\Hrp}$.
This constraint, which is Gauss' law for the magnetic field, is incorporated in~\eqref{sysin} via the reduction and restoration matrices~$\matr{F}_i^{(\mathrm{h})}$ and~$\matr{B}_i^{(\mathrm{h})}$, respectively, by dropping one unknown per object~\cite{Chew2009,JMMCTFSIE}.
Matrix~$\matr{F}_i^{(\mathrm{h})}$ drops one equation per object, while~$\matr{B}_i^{(\mathrm{h})}$ restores the dropped unknown in terms of the sum of the remaining unknowns~\cite{Chew2009}.
Importantly,~$\matr{F}_i^{(\mathrm{h})}$ and~$\matr{B}_i^{(\mathrm{h})}$ are defined independently for each object, even if objects are connected to each other directly or indirectly via circuit-like ports~\cite{JMMCTFSIE}, because of the absence of free magnetic charges flowing across object surfaces.
Therefore, in~\eqref{sysin},~$\matr{H}_i^{(\mathrm{n})}$ is a correspondingly reduced set of~${\nhat[i]'\cdot\Hrp}$ unknowns.

With this choice of integral equations, discretization, and constraint on~${\nhat[i]'\cdot\Hrp}$,~\eqref{sysin} is solvable down to exactly~$0\,$Hz, as confirmed in \secref{sec_res}.
Therefore,~$\matr{E}_i^{(\mathrm{t})}$ and~$\matr{H}_i^{(\mathrm{n})}$ in~\eqref{sysin} can be written as
\begin{align}
    \begin{bmatrix}
        \matr{E}_i^{(\mathrm{t})} \\
        \matr{H}_i^{(\mathrm{n})}
    \end{bmatrix} = \matr{X}_i\begin{bmatrix}
        \matr{H}_i^{(\mathrm{t})} \\
        \matr{E}_i^{(\mathrm{n})}
    \end{bmatrix},\label{sysinop}
\end{align}
where the immittance operator~$\matr{X}_i$ for object~$\Omega_{i}$, with~${i\in\{1,2,\ldots,N_{\mathrm{obj}}\}}$, is defined via~\eqref{sysin} as the solution to
\begin{multline}
    \begin{bmatrix}
        \nu\,\widetilde{\matr{K}}_{i}^{(\mathrm{r})} + \widetilde{\eps}_i\,\matr{L}_i^{(\mathrm{t})}
        &
        b_i^{(\mathrm{h})}\,\matr{D}_i^{\mathrm{T}}\matr{L}_i^{(\mathrm{s})}\matr{B}_i^{(\mathrm{h})}
        \\
        \matr{F}_i^{(\mathrm{h})}\matr{L}_i^{(\mathrm{s})}\matr{D}_i
        & -\widetilde{\mu}_i\,b_i^{(\mathrm{h})}\,\matr{F}_i^{(\mathrm{h})}\matr{L}_i^{(\mathrm{s})}\matr{B}_i^{(\mathrm{h})}
    \end{bmatrix}\matr{X}_i
    \\=
    -\begin{bmatrix}
        \nu\,\widetilde{\mu}_i\,\matr{L}_i^{(\mathrm{r})} + \widetilde{\matr{K}}_{i}^{(\mathrm{t})}
        &
        \nu\,b_i^{(\mathrm{e})}\,\matr{M}_{i}
        \\
        \matr{0}
        &
        \matr{0}
    \end{bmatrix},\label{sysinx}
\end{multline}
and is the discrete counterpart to the sought-after~$\mathcal{X}_i$ in~\eqref{xdef}.
The efficient solution of~\eqref{sysinx} for large problems will be discussed in \secref{sec_acc}.

\subsection{Exterior Region}\label{subsec_ext}

For~${i=0}$, the dual of~\eqref{sysin} is formulated using the same discretization scheme, but a different choice for the two integral equations: the first is a linear combination of~\eqref{tefie} and~\eqref{nmfie} tested with edge-length-normalized RWG functions, and the second is~\eqref{defie} tested with area-normalized triangle-wise constant functions, leading to
\begin{multline}
    \begin{bmatrix}
        \nu\,\widetilde{\matr{K}}_{0}^{(\mathrm{r})} - \widetilde{\mu}_0\,\matr{L}_0^{(\mathrm{t})}
        & \matr{D}_0^{\mathrm{T}}\matr{L}_0^{(\mathrm{s})}\matr{B}_0^{(\mathrm{e})}
        \\
        \matr{F}_0^{(\mathrm{e})}\matr{L}_0^{(\mathrm{s})}\matr{D}_0
        & \widetilde{\eps}_0\,\matr{F}_0^{(\mathrm{e})}\matr{L}_0^{(\mathrm{s})}\matr{B}_0^{(\mathrm{e})}
    \end{bmatrix}
    \begin{bmatrix}
        \matr{H}^{(\mathrm{t})} \\
        \matr{E}^{(\mathrm{n})}
    \end{bmatrix}
    \\+
    \begin{bmatrix}
        -\nu\,\widetilde{\eps}_0\,\matr{L}_0^{(\mathrm{r})} + \widetilde{\matr{K}}_{0}^{(\mathrm{t})}
        &
        \nu\matr{M}_{0}\matr{B}_0^{(\mathrm{h})}
        \\
        \matr{0}
        &
        \matr{0}
    \end{bmatrix}
    \begin{bmatrix}
        \matr{E}^{(\mathrm{t})} \\
        \matr{H}^{(\mathrm{n})}
    \end{bmatrix}
    \\=
    \begin{bmatrix}
        \matr{b}^{(\mathrm{t})} + \nu\matr{b}^{(\mathrm{r})} \\ \matr{0}
    \end{bmatrix}.\label{sysout}
\end{multline}
The entries of each matrix operator and the discretized right-hand side excitation vectors in~\eqref{sysout} are defined in the Appendix.
Reduction and restoration matrices~$\matr{F}_0^{(\mathrm{e})}$ and~$\matr{B}_0^{(\mathrm{e})}$ in~\eqref{sysout} are defined similarly to~$\matr{F}_i^{(\mathrm{h})}$ and~$\matr{B}_i^{(\mathrm{h})}$ in~\eqref{sysin}, respectively, and impose the Gauss' law constraint on~$\matr{E}^{(\mathrm{n})}$, which is a correspondingly reduced set of unknowns.
However,~$\matr{F}_0^{(\mathrm{e})}$ and~$\matr{B}_0^{(\mathrm{e})}$ must be defined \emph{per set of directly or indirectly connected objects}, because the associated free electric charges must be allowed to flow between connected objects~\cite{Chew2009,JMMCTFSIE}.
Matrix~$\matr{B}_0^{(\mathrm{h})}$ in~\eqref{sysout} is the block diagonal matrix with blocks~$\matr{B}_i^{(\mathrm{h})}$, for~${i\in\{1,2,\ldots,N_{\mathrm{obj}}\}}$.

\subsection{Final System of Equations}\label{subsec_fin}

The immittance operator for the entire structure is
\begin{align}
    \matr{X} = \bigoplus\limits_{i=1}^{N_{\mathrm{obj}}}\left(\matr{X}_i\right),\label{bigx}
\end{align}
assuming the appropriate object-wise layout of the unknown coefficient vectors.
In~\eqref{bigx},~$\bigoplus(\cdot)$ denotes a matrix direct sum of the term in the parentheses, indicating that the immittance operators for each object~$\Omega_i$,~${i\in\{1,2,\ldots,N_{\mathrm{obj}}\}}$, are arranged into a single block diagonal matrix.
Then, using~\eqref{bigx} in~\eqref{sysout} to eliminate~$\matr{E}^{(\mathrm{t})}$ and~$\matr{H}^{(\mathrm{n})}$, and making use of the interface conditions in \secref{subsec_bc}, the final system of equations is
\begin{multline}
    \left(%
    \begin{bmatrix}
        \nu\,\widetilde{\matr{K}}_{0}^{(\mathrm{r})} - \widetilde{\mu}_0\,\matr{L}_0^{(\mathrm{t})}
        & \matr{D}_0^{\mathrm{T}}\matr{L}_0^{(\mathrm{s})}\matr{B}_0^{(\mathrm{e})}
        \\
        \matr{F}_0^{(\mathrm{e})}\matr{L}_0^{(\mathrm{s})}\matr{D}_0
        & \widetilde{\eps}_0\,\matr{F}_0^{(\mathrm{e})}\matr{L}_0^{(\mathrm{s})}\matr{B}_0^{(\mathrm{e})}
    \end{bmatrix}
    \right.\\ \left.+
    \begin{bmatrix}
        -\nu\,\widetilde{\eps}_0\,\matr{L}_0^{(\mathrm{r})} + \widetilde{\matr{K}}_{0}^{(\mathrm{t})}
        &
        \nu\matr{M}_{0}\matr{B}_0^{(\mathrm{h})}
        \\
        \matr{0}
        &
        \matr{0}
    \end{bmatrix}
    \matr{X}
    \right)
    \begin{bmatrix}
        \matr{H}^{(\mathrm{t})} \\
        \matr{E}^{(\mathrm{n})}
    \end{bmatrix}
    \\=
    \begin{bmatrix}
        \matr{b}^{(\mathrm{t})} + \nu\matr{b}^{(\mathrm{r})} \\ \matr{0}
    \end{bmatrix}.\label{sys}
\end{multline}
The form of~\eqref{sys} resembles that of the AGIBC~\cite{QianChew2009} and the SLIM~\cite{AWPLSLIM} formulations: the first term in parentheses in~\eqref{sys} models the exterior region, and would be the only term if the objects were perfect electric conductors.
The second term in parentheses in~\eqref{sys} can be seen as a correction term that models the region interior to each object, when those objects are not perfect conductors.
However, the key distinction compared to~\cite{QianChew2009} and~\cite{AWPLSLIM} is that the proposed formulation in~\eqref{sys} is accurate and stable from~$0\,$Hz to the full-wave regime, as demonstrated in \secref{sec_res}.
At the same time,~\eqref{sys} has the same number of unknowns as the formulations in~\cite{QianChew2009} and~\cite{AWPLSLIM}, unlike the frequency-stable F-TENH/NETH formulation in~\cite{JMMCTFSIE}, which has about twice as many unknowns.
Therefore, the proposed method retains the frequency-stability of the one in~\cite{JMMCTFSIE}, but with improved efficiency, as confirmed in \secref{sec_res}.
The implementation of~\eqref{sys} with matrix compression is discussed next in \secref{sec_acc}, to obtain approximately log-linear time and memory complexity.

\section{Matrix Compression and Solution}\label{sec_acc}

Acceleration or matrix compression algorithms are usually necessary in boundary element methods to provide a log-linear, or at least sub-quadratic, time and memory complexity, which would otherwise scale as the square of the number of unknowns at best.
In this work, the following compression and fast solution strategy is proposed for solving~\eqref{sys} efficiently:
\begin{itemize}
    \item Each integral equation operator is computed in compressed form as a hierarchical ($\mathcal{H}$)-matrix~\cite{Hackbusch1999,BookHackbusch2014}.
    \item A block hierarchical lower-upper (\hlu{}) factorization~\cite{BookHackbusch2014} is used to factorize the system in~\eqref{sysinx}, for~${i\in\{1,2,\ldots,N_{\mathrm{obj}}\}}$.
    \item An iterative solver is used to solve~\eqref{sys}, where the factors associated with each~$\matr{X}_i$ are applied in a hierarchical backward-forward substitution at each iteration.
\end{itemize}
Each step is described in detail in Sections~\ref{sec_accass}--\ref{sec_accsol}.
Note that the proposed strategy can be adapted to other compression formats, such as~$\mathcal{H}^2$~\cite{Hackbusch2000}, HODLR~\cite{Hackbusch2004,AmbikasaranDarve2013}, and HSS~\cite{Pals2006} matrices, that offer a sub-quadratic LU factorization.

\subsection{Operator Assembly}\label{sec_accass}

Each integral operator matrix in~\eqref{sysinx} and~\eqref{sys} is assembled directly in compressed form using the~\hmat{} format.
A detailed mathematical description of the structure and construction of~\hmats{} can be found in~\cite{Hackbusch1999,BookHackbusch2014}; a brief summary is provided below, with a focus on the aspects that are specific to the proposed formulation.

The~\hmat{} format relies on a hierarchical subdivision of mesh elements into a binary tree, referred to as a cluster tree.
In this work, a balanced geometric subdivision of mesh elements is used: the geometry is enclosed in an axis-aligned bounding box, which is recursively subdivided along its longest axis such that each partition contains an approximately equal number of mesh elements.
The subdivision stops once a partition contains fewer than~${2N_{\mathrm{leaf}}}$ elements, and such a partition is called a leaf, where~$N_{\mathrm{leaf}}$ is a prescribed leaf size.
Each partition is called a cluster, and together, these hierarchical clusters constitute a binary cluster tree.
Electromagnetic interactions are grouped into pairs of interacting source and observation clusters, and each pair forms a block cluster.
The hierarchy of block clusters constitutes a block cluster tree, where each node corresponds to a block of the matrix being compressed.

The compressed representation of the matrix is obtained by recognizing that pairs of clusters that are sufficiently far apart permit an approximate low-rank representation of the corresponding matrix block.
Blocks that can be approximated accurately in this way are referred to as ``admissible'', and are labeled based on an admissibility criterion.
In this work, the strong admissibility criterion is used, where a pair of clusters is considered admissible if
\begin{align}
    \eta\,\mathrm{dist}\left(\mathcal{C}_{\mathrm{src}}, \mathcal{C}_{\mathrm{obs}}\right) \geq \min\left(\mathrm{diam}\left(\mathcal{C}_{\mathrm{src}}\right), \mathrm{diam}\left(\mathcal{C}_{\mathrm{obs}}\right)\right),\label{adm}
\end{align}
where~$\eta$ is a dimensionless scale factor, typically~$\mathcal{O}(1)$, and~$\mathcal{C}$ denotes a cluster of mesh elements, with the~$\mathrm{src}$ and~$\mathrm{obs}$ subscripts indicating a source and observation cluster, respectively.
The distance~$\mathrm{dist}(\cdot)$ between two clusters is approximated as the Euclidean distance between the associated bounding boxes, and the diameter~$\mathrm{diam}(\cdot)$ of a cluster is approximated as the diameter of its bounding box~\cite{DJiao2013}.
In addition to~\eqref{adm}, the extended admissibility condition of~\cite{Mao2018} is used, whereby a parent block is considered admissible if each of its child blocks is admissible per~\eqref{adm}, even if the condition in~\eqref{adm} is not met for the parent block itself.
Inadmissible matrix blocks are not compressed, and instead are computed and stored in dense format, with dense blocks typically constituting a small fraction of the full matrix.

To obtain a compressed representation of admissible matrix blocks, the adaptive cross approximation~(ACA)~\cite{Bebendorf2000,ZhaoJFLee2005} can be used to approximate matrix block~${\matr{Z}\in N_{\mathrm{obs}}\times N_{\mathrm{src}}}$ as
\begin{align}
    \matr{Z} \approx \matr{U}\matr{V}^{\mathrm{T}},\label{aca}
\end{align}
where~${\matr{U} \in \mathbb{C}^{N_{\mathrm{obs}}\times r}}$ and~${\matr{V} \in \mathbb{C}^{N_{\mathrm{src}}\times r}}$, with~$N_{\mathrm{src}}$ and~$N_{\mathrm{obs}}$ denoting the number of source and observation degrees of freedom in the block~$\matr{Z}$, and~${r \ll N_{\mathrm{obs}}, N_{\mathrm{src}}}$.
In this work, the ACA+ algorithm~\cite{Grasedyck2005} with QR-SVD recompression~\cite{BebendorfRjasanow2003,BebendorfKunis2009} was used to improve compression quality and efficiency.
The ACA is an iterative procedure which involves picking pivot rows and columns based on a running estimate of the residual between the approximation and the ground truth.
The ACA+ makes pivot selection more robust by keeping track of a reference row and column of the residual estimate, whose largest entries inform pivot selection at subsequent iterations.

Equations~\eqref{sysinx} and~\eqref{sys} both involve rectangular operator matrices that mix degrees of freedom associated with mesh triangles and mesh edges, such as~$\matr{M}_i$, for which independent edge-based and triangle-based cluster trees are generated.
In both cases, a cluster's bounding box is defined such that it encloses every mesh vertex of all elements (triangles or edges) belonging to that cluster.
A node of the associated block cluster tree then denotes interactions between a triangle-based cluster and an edge-based cluster, and the admissibility criterion~\eqref{adm} with the extension in~\cite{Mao2018} continues to apply.

A key feature of the proposed formulation is that the immittance operator is defined independently for each object.
Therefore, the generation of cluster and block trees, and the subsequent compressed matrix assembly for the operators in~\eqref{sysinx}, are performed independently for the mesh of each object, and therefore, independently for regions composed of different materials.
This has important advantages compared to existing methods in the literature~\cite{Xie2022,Ren2024}:
\begin{itemize}
    \item In existing combined-field formulations that involve adding together integral operators for different materials~\cite{Oijala2005}, the formatted addition of the associated~\hmats{} would require a shared block cluster tree between the exterior and interior region operators~\cite{BookHackbusch2014}.
    Therefore, a global clustering strategy may be required, where a single cluster and block tree is used for the interior region of all objects.
    Since the interior-region matrix operators for an object are non-zero only for interactions local to that object, the global matrix containing all interior region interactions will be zero for all source and observation degrees of freedom that are not associated with the same object.
    When attempting to compress blocks of this matrix with ACA using a shared block cluster tree, extra care and bookkeeping are required to maintain accuracy~\cite{Ren2024}.
    In the proposed formulation, the compression is performed independently for the operators associated with each object, based on independent cluster trees.
    This avoids any complications related to partially zero matrix blocks, and exactly preserves the property that interior region interactions are local to objects.
    \item 
    When shared clustering is used to capture interior region interactions, as described in the previous point, individual matrix blocks processed by ACA may involve terms associated with objects of different material properties.
    These blocks may thus have entries of varying orders of magnitude, potentially leading to premature termination of the ACA iterations, and thus an inaccurate compressed representation.
    This would require either using more advanced compression strategies, or further dividing such matrix blocks into smaller blocks local to each object, each one compressed separately.
    In the proposed approach, because of its object-wise construction for interior-region interactions, this issue is avoided.
    \item For structures with many objects, the proposed object-wise independent compression strategy allows for the parallel assembly of matrix operators for different objects, on multicore or multinode compute systems.
    As ACA is an iterative algorithm, exploiting parallelism across objects might be more efficient than restricting parallelism to within a single large matrix compression step, depending on the type of problem.
    However, this point is left as a potential future opportunity, and its careful study is beyond the scope of this article.
\end{itemize}

\subsection{Immittance Operator Factorization}\label{sec_acchlu}

To obtain the immittance operator~$\matr{X}_i$ for~$i\in\{1,2,\ldots,N_{\mathrm{obj}}\}$, the system in~\eqref{sysinx} must be solved for~$\matr{X}_i$,
\begin{multline}
    \matr{X}_i
    =
    -\begin{bmatrix}
        \nu\,\widetilde{\matr{K}}_{i}^{(\mathrm{r})} + \widetilde{\eps}_i\,\matr{L}_i^{(\mathrm{t})}
        &
        b_i^{(\mathrm{h})}\,\matr{D}_i^{\mathrm{T}}\matr{L}_i^{(\mathrm{s})}\matr{B}_i^{(\mathrm{h})}
        \\
        \matr{F}_i^{(\mathrm{h})}\matr{L}_i^{(\mathrm{s})}\matr{D}_i
        & -\widetilde{\mu}_i\,b_i^{(\mathrm{h})}\,\matr{F}_i^{(\mathrm{h})}\matr{L}_i^{(\mathrm{s})}\matr{B}_i^{(\mathrm{h})}
    \end{bmatrix}^{-1}
    \cdot\\
    \begin{bmatrix}
        \nu\,\widetilde{\mu}_i\,\matr{L}_i^{(\mathrm{r})} + \widetilde{\matr{K}}_{i}^{(\mathrm{t})}
        &
        \nu\,b_i^{(\mathrm{e})}\,\matr{M}_{i}
        \\
        \matr{0}
        &
        \matr{0}
    \end{bmatrix} = \matr{P}_i^{-1}\matr{Q}_i,\label{sysinxinv}
\end{multline}
where symbols~$\matr{P}_i$ and~$\matr{Q}_i$ have been introduced for convenience.
Since~$\matr{X}_i$ is eventually used in the context of an iterative solver, we require only its product with a vector, so the right-hand side of~\eqref{sysinxinv} need not be assembled explicitly.
Instead,~$\matr{P}_i$ is factorized once in advance, and the factors are later applied with forward and backward substitution during the iterative solve step, as described in \secref{sec_accsol}.
The goal of this section is to describe the efficient factorization of~$\matr{P}_i$.

It is desirable to leverage the~\hmat{} representation discussed in \secref{sec_accass}, and use a fast block~\hlu{} factorization of~$\matr{P}_i$.
However, this requires each block of~$\matr{P}_i$ to be stored in~\hmat{} format.
The compressed assembly procedure described in \secref{sec_accass} provides \emph{individual matrix operators} in an~\hmat{} format, but does not address their multiplication with the sparse matrices~$\matr{D}_i$,~$\matr{F}_i^{\mathrm{(h)}}$, and~$\matr{B}_i^{\mathrm{(h)}}$ that appear in~$\matr{P}_i$.
Therefore, terms involving these sparse matrices, and the subsequent block factorization, are handled as follows:

\subsubsection{Discrete divergence sparse matrix}\label{sec_acchlu_d}

The entries of~${\matr{D}_i^{\mathrm{T}}\matr{L}_i^{(\mathrm{s})}}$ and~${\matr{L}_i^{(\mathrm{s})}\matr{D}_i}$ in~\eqref{sysinxinv} can be written directly as
\begin{align}
    \matr{L}_{i,(l,m)}^{(\mathrm{ds})} &\triangleq
    \sum_q^{N_{\mathrm{face},i}} \matr{D}_{i,(l,q)}^{\mathrm{T}}\matr{L}_{i,(q,m)}^{(\mathrm{s})},\label{opDLs}\\
    \matr{L}_{i,(l,m)}^{(\mathrm{sd})} &\triangleq 
    \sum_q^{N_{\mathrm{face},i}} \matr{L}_{i,(l,q)}^{(\mathrm{s})}\matr{D}_{i,(q,m)},\label{opLsD}
\end{align}
respectively.
In~\eqref{opDLs} and~\eqref{opLsD},~${\matr{L}_{i}^{(\mathrm{ds})}\in\mathbb{C}^{N_{\mathrm{edge},i}\times N_{\mathrm{face},i}}}$, and~${\matr{L}_{i}^{(\mathrm{sd})}\in\mathbb{C}^{N_{\mathrm{face},i}\times N_{\mathrm{edge},i}}}$.
Since~$\matr{D}_i$ has only two non-zero entries per column, the summations in~\eqref{opDLs} and~\eqref{opLsD} are actually over only two terms, corresponding to the two triangles associated with a given edge.
In fact, computing the entries of~$\matr{L}_{i}^{(\mathrm{ds})}$ and~$\matr{L}_{i}^{(\mathrm{sd})}$ in this way is a standard part of integral equation codes where the hypersingular operator is computed directly~\cite{BookChew2009}.
Then, the rectangular matrices~$\matr{L}_{i}^{(\mathrm{ds})}$ and~$\matr{L}_{i}^{(\mathrm{sd})}$ can be assembled directly in compressed form using the same mixed-degree-of-freedom clustering as~$\matr{M}_i$.

\subsubsection{Unknown reduction sparse matrix}\label{sec_acchlu_f}

Entries of~${\matr{F}_i^{(\mathrm{h})}\in\mathbb{C}^{(N_{\mathrm{face},i}-1)\times N_{\mathrm{face},i}}}$ can be defined, in accordance with the discussion in \secref{subsec_int}, as
\begin{align}
    \matr{F}_{i,(l,m)}^{(\mathrm{h})} = 
    \begin{cases}
        1 & l = m,\\
        0 & m = N_{\mathrm{face},i},\\
        0 & \mathrm{otherwise},
    \end{cases}\label{opF}
\end{align}
for row~$l$ and column~$m$, where a~$1$-based indexing is used, and it is assumed for simplicity, without loss of generality, that the~${\nhat[i]'\cdot\Hrp}$ unknown to be dropped is the last one in the associated vector of coefficients.
Therefore,~${\matr{F}_i^{(\mathrm{h})}\matr{L}_i^{(\mathrm{s})}}$ in~\eqref{sysinxinv} implies pruning the last row of~$\matr{L}_i^{(\mathrm{s})}$.
After obtaining an~\hmat{} representation of~$\matr{L}_i^{(\mathrm{s})}$, applying~$\matr{F}_i^{(\mathrm{h})}$ is straightforward:
for all admissible or inadmissible matrix blocks composing~$\matr{L}_i^{(\mathrm{s})}$ that contain the dropped row, the corresponding row of the matrix block is deleted.
For an admissible block represented as~$\matr{U}\matr{V}^{\mathrm{T}}$, this amounts to deleting the corresponding row of~$\matr{U}$.
For an inadmissible block that is stored in dense format, the corresponding row is deleted.
Therefore,~${\matr{F}_i^{(\mathrm{h})}\matr{L}_i^{(\mathrm{s})}}$ in~\eqref{sysinxinv} is a rectangular~\hmat{}.
From an implementation perspective, the operations for both admissible and inadmissible blocks involve a memory reallocation and move of the retained rows of~$\matr{U}$ or the dense block, respectively, which does not impact the overall computational complexity.

\subsubsection{Unknown restoration sparse matrix}\label{sec_acchlu_b}

Entries of~${\matr{B}_i^{(\mathrm{h})}\in\mathbb{C}^{N_{\mathrm{face},i}\times (N_{\mathrm{face},i}-1)}}$ can be defined, also in accordance with the discussion in \secref{subsec_int}, as
\begin{align}
    \matr{B}_{i,(l,m)}^{(\mathrm{h})} = 
    \begin{cases}
        1 & l = m,\\
        -1 & l = N_{\mathrm{face},i},\\
        0 & \mathrm{otherwise},
    \end{cases}\label{opB}
\end{align}
for row~$l$ and column~$m$.
The term~${\matr{L}_i^{(\mathrm{s})}\matr{B}_i^{(\mathrm{h})}}$ in~\eqref{sysinxinv} is therefore responsible for restoring the dropped~${\nhat[i]'\cdot\Hrp}$ unknown as the negative of the sum of the retained~${\nhat[i]'\cdot\Hrp}$ unknowns.
Matrix~$\matr{B}_i^{(\mathrm{h})}$ can be written in an alternative form as
\begin{align}
    \matr{B}_i^{(\mathrm{h})} = (\matr{F}_i^{(\mathrm{h})})^{\mathrm{T}} - \matr{e}_{\mathrm{d}}\matr{\mathbb{1}}^{\mathrm{T}},\label{opB2}
\end{align}
where entries of column vectors~${\matr{e}_{\mathrm{d}}\in N_{\mathrm{face},i}\times 1}$ and~${\matr{\mathbb{1}}\in (N_{\mathrm{face},i}-1)\times 1}$ are defined as
\begin{align}
    \matr{e}_{\mathrm{d},(l)} &=
    \begin{cases}
        1 & l = N_{\mathrm{face},i},\\
        0 & \mathrm{otherwise},
    \end{cases}\\
    \matr{\mathbb{1}}_{(l)} &= 1,
\end{align}
respectively, for row~$l$.
Therefore,~$\matr{B}_i^{(\mathrm{h})}$ can be applied to the~\hmat{} representation of~$\matr{L}_i^{(\mathrm{s})}$ as
\begin{align}
    \matr{L}_i^{(\mathrm{s})}\matr{B}_i^{(\mathrm{h})} = \matr{L}_i^{(\mathrm{s})}(\matr{F}_i^{(\mathrm{h})})^{\mathrm{T}} - (\matr{L}_i^{(\mathrm{s})}\matr{e}_{\mathrm{d}})\matr{\mathbb{1}}^{\mathrm{T}},\label{LB}
\end{align}
where~${\matr{L}_i^{(\mathrm{s})}\matr{e}_{\mathrm{d}}}$ is a single matrix-vector product.
Therefore, the first term on the right-hand side of~\eqref{LB} is computed as in the previous point, by pruning the last column of~$\matr{L}_i^{(\mathrm{s})}$.
The second term on the right-hand side of~\eqref{LB} is a rank-one correction to the~\hmat{}~${\matr{L}_i^{(\mathrm{s})}(\matr{F}_i^{(\mathrm{h})})^{\mathrm{T}}}$, which can be applied using formatted addition, a standard operation in~\hmat{} arithmetic~\cite{BookHackbusch2014}.
Therefore,~${\matr{L}_i^{(\mathrm{s})}\matr{B}_i^{(\mathrm{h})}}$ in~\eqref{sysinxinv} can be represented as a rectangular~\hmat{}.
Since any submatrix block of~${(\matr{L}_i^{(\mathrm{s})}\matr{e}_{\mathrm{d}})\matr{\mathbb{1}}^{\mathrm{T}}}$ has a rank of at most one, this correction term can be applied block by block.
For inadmissible blocks, the correction can be subtracted directly.
For an admissible block~$\matr{U}\matr{V}^{\mathrm{T}}$, the corresponding segments of~${(-\matr{L}_i^{(\mathrm{s})}\matr{e}_{\mathrm{d}})}$ and~$\matr{\mathbb{1}}$ are appended as an extra column to~$\matr{U}$ and~$\matr{V}$, respectively, followed by a recompression at machine precision~\cite{BookHackbusch2014}.
The block structure of the~\hmat{} is therefore unchanged, and the rank of each admissible block grows by at most one.

\subsubsection{Block factorization}\label{sec_acchlu_hlu}

With~$\matr{P}_i$ in \eqref{sysinxinv} now written as a block~\hmat{}, a standard block~\hlu{}~\cite{BookHackbusch2014} can be applied with a time complexity of~$\mathcal{O}(N_i\log^2N_i)$ and a memory complexity of~$\mathcal{O}(N_i\log N_i)$ up to moderate electrical sizes, where~${N_i=N_{\mathrm{face},i} + N_{\mathrm{edge},i}}$~\cite{BookHackbusch2014}.
Denoting the~${2\times2}$ blocks of~$\matr{P}_i$ in~\eqref{sysinxinv} by~$\matr{P}_{i,lm}$, the block~\hlu{} factorization computes
\begin{align}
    \matr{P}_i =
    \begin{bmatrix}
        \mathcal{L}_{i,11} & \matr{0}\\
        \mathcal{L}_{i,21} & \mathcal{L}_{i,22}
    \end{bmatrix}
    \begin{bmatrix}
        \mathcal{U}_{i,11} & \mathcal{U}_{i,12}\\
        \matr{0} & \mathcal{U}_{i,22}
    \end{bmatrix},\label{blockhlu}
\end{align}
where~$\mathcal{L}_{i,11}$ and~$\mathcal{L}_{i,22}$ are lower-triangular,~$\mathcal{U}_{i,11}$ and~$\mathcal{U}_{i,22}$ are upper-triangular, and each block is an~\hmat{} with the same structure as the corresponding block of~$\matr{P}_i$.
Equating both sides of~\eqref{blockhlu} blockwise,~$\mathcal{L}_{i,lm}$ and~$\mathcal{U}_{i,lm}$ are computed as follows:
\begin{enumerate}
    \item
    Block~$\matr{P}_{i,11}$ is factorized with a standard~\hlu{}~\cite{BookHackbusch2014},
    \begin{align}
        \mathcal{L}_{i,11}\,\mathcal{U}_{i,11} = \matr{P}_{i,11}.\label{blockhlu1}
    \end{align}
    \item
    Given~$\mathcal{L}_{i,11}$ from~\eqref{blockhlu1},~$\mathcal{U}_{i,12}$ is obtained from
    \begin{align}
        \mathcal{L}_{i,11}\,\mathcal{U}_{i,12} = \matr{P}_{i,12}\label{blockhlu2}
    \end{align}
    with a lower-triangular~\hmat{} solve via forward substitution, whose right-hand side is~$\matr{P}_{i,12}$~\cite{BookHackbusch2014}.
    \item
    Similarly, given~$\mathcal{U}_{i,11}$, block~$\mathcal{L}_{i,21}$ is obtained from
    \begin{align}
        \mathcal{L}_{i,21}\,\mathcal{U}_{i,11} = \matr{P}_{i,21}\label{blockhlu3}
    \end{align}
    by an upper-triangular~\hmat{} solve from the right.
    \item
    Finally, the Schur complement of~$\matr{P}_{i,11}$ is formed,
    \begin{align}
        \matr{S}_i &= \matr{P}_{i,22} - \mathcal{L}_{i,21}\,\mathcal{U}_{i,12} \\&= \matr{P}_{i,22} - \matr{P}_{i,21}\,\matr{P}_{i,11}^{-1}\,\matr{P}_{i,12},\label{blockhlu4S}
    \end{align}
    using formatted~\hmat{} multiplication and addition~\cite{BookHackbusch2014}.
    The~\hmat{}~$\matr{S}_i$ is then factorized with a standard~\hlu{} as
    \begin{align}
        \mathcal{L}_{i,22}\,\mathcal{U}_{i,22} = \matr{S}_i.\label{blockhlu4}
    \end{align}
\end{enumerate}

All~\hmat{} operations above are truncated to a prescribed~\hlu{} tolerance, which keeps the ranks of admissible blocks bounded and preserves the block structure of each factor~\cite{BookHackbusch2014}.
Note that the top-left block~$\matr{P}_{i,11}$ must be used as the pivot and eliminated first.
This is because when~${\omega=0}$, the coefficient of~$\matr{P}_{i,22}$ in~\eqref{sysinxinv} vanishes so that~${\matr{P}_{i,22}=\matr{0}}$.
With~$\matr{P}_{i,11}$ as the pivot instead,~\eqref{blockhlu4S} reduces to~${\matr{S}_i = -\matr{P}_{i,21}\,\matr{P}_{i,11}^{-1}\,\matr{P}_{i,12}}$ when~${\omega=0}$, so the factorization in~\eqref{blockhlu4} remains stable.
In-house implementations of~\hlu{} and of the block factorization were used in this work, but each of the four steps above corresponds directly to functionality available in standard~\hmat{} libraries, such as HLIBpro~\cite{hlibpro}.

\subsection{Final System Solution}\label{sec_accsol}

Equation~\eqref{sys} is solved with an iterative solver such as~GMRES~\cite{gmres}, which requires computing the product of the system matrix in~\eqref{sys} with a vector.
Assuming that~$\matr{x}^q$ represents the approximate solution to~\eqref{sys} at the~$q^{\mathrm{th}}$ iteration, the matrix-vector product is computed as follows:
\begin{enumerate}
    \item The immittance operator~$\matr{X}$ is applied as
    \begin{align}
        \matr{x}_1 = \matr{X}\matr{x}^q
        = \bigoplus\limits_{i=1}^{N_{\mathrm{obj}}}\left(\matr{P}_i^{-1}\matr{Q}_i\matr{x}_i^q\right),
    \end{align}
    where~$\matr{x}_i^q$ is the subvector of~$\matr{x}^q$ corresponding to object~$i$, and~$\bigoplus(\cdot)$ for column vectors denotes vertical concatenation.
    The term~$\matr{Q}_i\matr{x}_i^q$ is computed first with blockwise~\hmat{}-vector products, and~${\matr{P}_i^{-1}(\matr{Q}_i\matr{x}_i^q)}$ is then computed using forward and backward substitution of the factors of~$\matr{P}_i$ from \secref{sec_acchlu}.

    \item The correction term that models the interior regions' coupling to the exterior region is computed with blockwise matrix-vector products as
    \begin{align}
        \matr{x}_2 &= \begin{bmatrix}
            -\nu\,\widetilde{\eps}_0\,\matr{L}_0^{(\mathrm{r})} + \widetilde{\matr{K}}_{0}^{(\mathrm{t})}
            &
            \nu\matr{M}_{0}\matr{B}_0^{(\mathrm{h})}
            \\
            \matr{0}
            &
            \matr{0}
        \end{bmatrix}
        \matr{x}_1.
    \end{align}

    \item The exterior region contribution is computed with blockwise matrix-vector products as
    \begin{align}
        &\matr{x}_3 = \begin{bmatrix}
            \nu\,\widetilde{\matr{K}}_{0}^{(\mathrm{r})} - \widetilde{\mu}_0\,\matr{L}_0^{(\mathrm{t})}
            & \matr{D}_0^{\mathrm{T}}\matr{L}_0^{(\mathrm{s})}\matr{B}_0^{(\mathrm{e})}
            \\
            \matr{F}_0^{(\mathrm{e})}\matr{L}_0^{(\mathrm{s})}\matr{D}_0
            & \widetilde{\eps}_0\,\matr{F}_0^{(\mathrm{e})}\matr{L}_0^{(\mathrm{s})}\matr{B}_0^{(\mathrm{e})}
        \end{bmatrix}
        \matr{x}^q.
    \end{align}

    \item The final matrix-vector product is simply~${\matr{x}_2 + \matr{x}_3}$.
\end{enumerate}

In the second and third steps above, one never needs to compute the product of an integral operator matrix with the sparse matrices~$\matr{D}_0$,~$\matr{F}_0^{\mathrm{(e)}}$, and~$\matr{B}_0^{\mathrm{(e)}}$, because each individual matrix block is first applied to the multiplying vector on the right.
Therefore, each integral operator matrix,~$\matr{L}_0^{(\mathrm{t})}$,~$\matr{L}_0^{(\mathrm{r})}$,~$\matr{L}_0^{(\mathrm{s})}$,~$\widetilde{\matr{K}}_{0}^{(\mathrm{t})}$,~$\widetilde{\matr{K}}_{0}^{(\mathrm{r})}$, and~$\matr{M}_{0}$ is assembled as a compressed~\hmat{} similarly to the interior-region matrices in \secref{sec_accass}, and applied to the vector on the right using a standard~\hmat{}-vector product~\cite{BookHackbusch2014}.

\subsection{Preconditioning}\label{sec_pc}

To speed up the convergence of iterative solvers for solving~\eqref{sys} as described in \secref{sec_accsol}, the following sparse right-preconditioner is used,
\begin{align}
    \widetilde{\matr{A}} = \begin{bmatrix}
        \mathrm{diag}\left(\nu\,\widetilde{\matr{K}}_{0}^{(\mathrm{r})} - \widetilde{\mu}_0\,\matr{L}_0^{(\mathrm{t})}\right)
        & \matr{D}_0^{\mathrm{T}}\mathrm{diag}\left(\matr{L}_0^{(\mathrm{s})}\right)\matr{B}_0^{(\mathrm{e})}
        \\
        \matr{F}_0^{(\mathrm{e})}\mathrm{diag}\left(\matr{L}_0^{(\mathrm{s})}\right)\matr{D}_0
        & \widetilde{\eps}_0\,\matr{F}_0^{(\mathrm{e})}\mathrm{diag}\left(\matr{L}_0^{(\mathrm{s})}\right)\matr{B}_0^{(\mathrm{e})}
    \end{bmatrix},\label{pc}
\end{align}
where~$\mathrm{diag}(\cdot)$ denotes keeping only diagonal entries of the matrix in parentheses.
Using the Schur complement-based technique in~\cite{Chew2009}, the preconditioner~$\widetilde{\matr{A}}$ can be factorized efficiently with sparse direct solvers~\cite{mumps,superlu}.
Note that~$\widetilde{\matr{A}}$ contains only the contributions from exterior-region operators.
Incorporating contributions from interior-region operators into the preconditioner may further improve the convergence of iterative solvers, but this investigation is left for future work.
The performance of~$\widetilde{\matr{A}}$ for low to moderate frequencies is adequate, as shown in the numerical examples in \secref{sec_res}.
The size of~$\widetilde{\matr{A}}$ is approximately half that of the preconditioner used in~\cite{JMMCTFSIE}, which makes it significantly faster to factorize and apply at each iteration, with a smaller memory footprint, as demonstrated in \secref{sec_res_int}.

\section{Numerical Results}\label{sec_res}

Numerical examples spanning static, quasistatic, and full-wave electromagnetic regimes are presented in this section.
All simulations were performed multithreaded with 16 threads, with a simple shared-memory parallelization that was not optimized, since parallel computation is not the focus of this article.
Unless otherwise mentioned, the ACA convergence and recompression tolerance,~\hlu{} recompression tolerance, and GMRES convergence relative residual were all set to~$10^{-4}$.
For constructing the~\hmats{}, a leaf size of~\num{32} and~$\eta$ of~\num{4} were used in all examples.
For computing integral operators for highly conductive materials, the line integration technique in~\cite{Chew2007} and its modification in~\cite{Chew2017} were used.

The proposed method and the one in~\cite{JMMCTFSIE} were implemented using the open-source boundary element framework, \href{https://github.com/shash-sharma/OpenBEM}{OpenBEM}~\cite{OpenBEM}.
The geometries and associated meshes were generated using the Python interface to Gmsh~\cite{gmsh}, which was also used for visualizing surface current distributions.
Scattering parameters were post-processed using the {scikit-rf} package~\cite{skrf_mini} in Python.
CAD files have been provided as supplementary material for the examples in \secref{sec_res_int} and \secref{sec_res_arr}, so detailed geometric dimensions have not been provided for them in the text, to save space.

To ensure fair performance comparisons between the proposed method and the F-TENH/NETH formulation in~\cite{JMMCTFSIE}, the~\hmat{} compression strategy described in \secref{sec_accass} was also applied for the exterior and interior region matrix operators in the F-TENH/NETH implementation.
The sparse matrix operations in \secref{sec_acchlu} were not needed for the F-TENH/NETH method because it does not require interior-region~\hlu{} factorizations.

\subsection{DC Capacitance, Inductance, and Resistance Extraction}\label{sec_res_clr}

\begin{figure}[t]
  \centering
  \includegraphics[width=\linewidth]{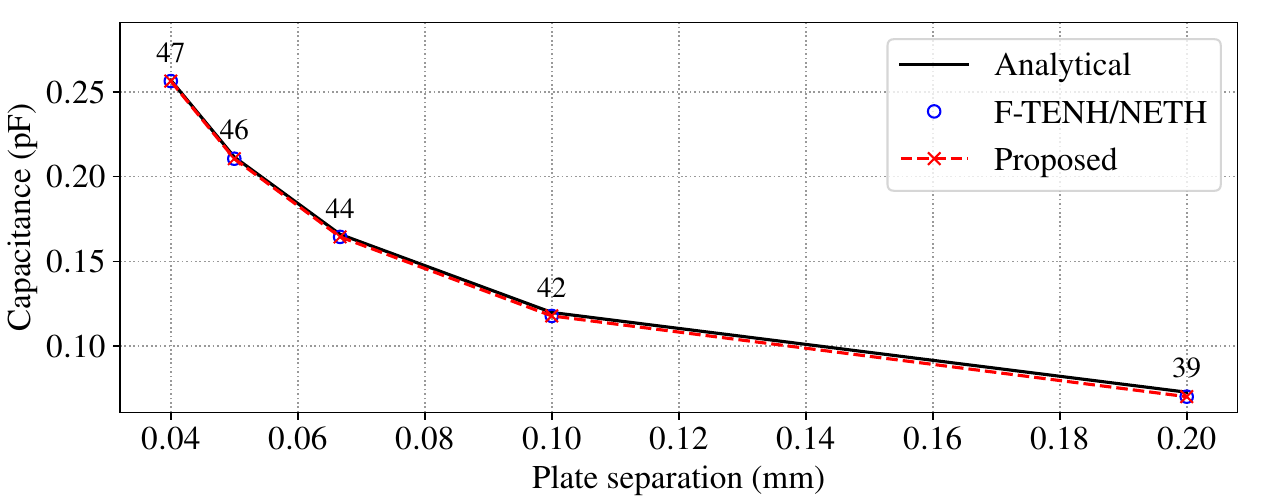}
  \caption{Capacitance as a function of plate separation for the parallel plate capacitor in \secref{sec_res_clr} at~\qty{0}{\hertz}.}\label{fig_pp}
\end{figure}

\begin{figure}[t]
  \centering
  \includegraphics[width=\linewidth]{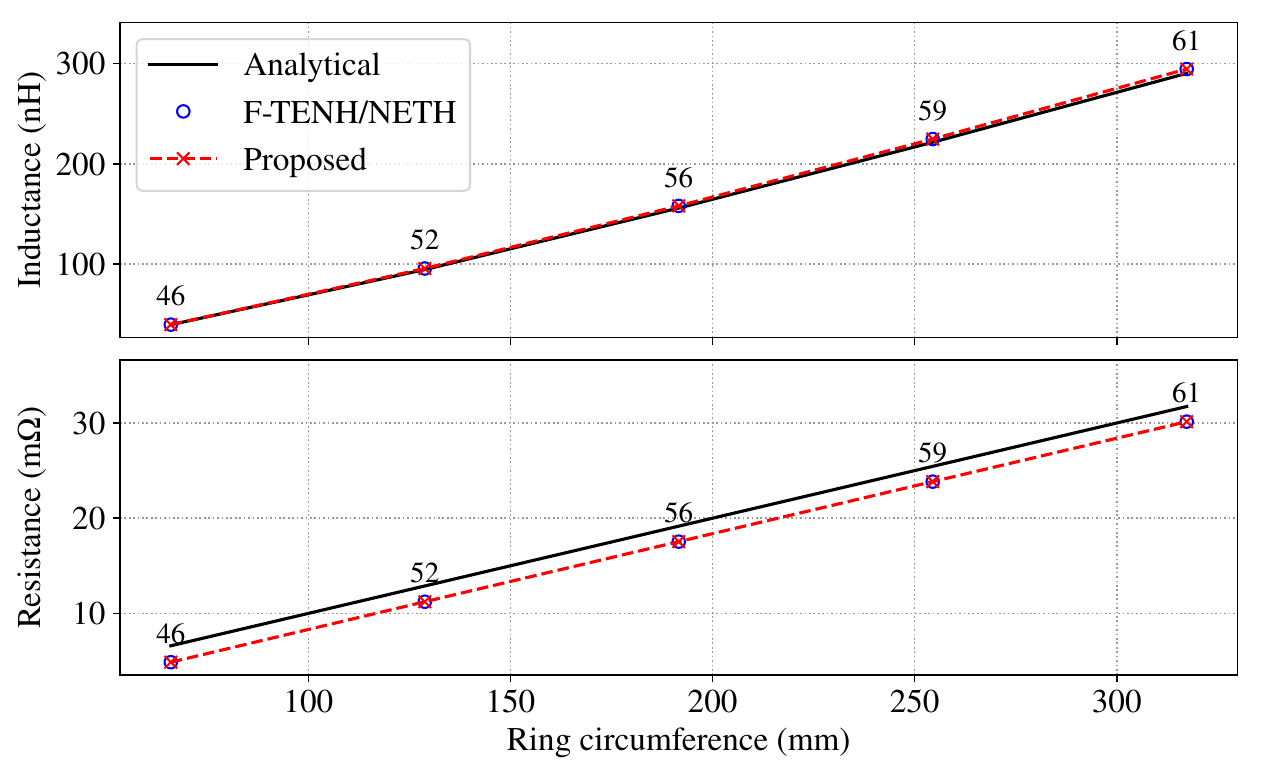}
  \caption{Inductance (upper panel) and resistance (lower panel) as a function of circumference for the ring in \secref{sec_res_clr} at~\qty{0}{\hertz}.}\label{fig_ring}
\end{figure}

To validate the accuracy of the proposed method in the electrostatic and magnetostatic limits, a parallel plate capacitor and a single-turn coil, both with an electrical conductivity of~\qty{e7}{\siemens\per\meter}, were simulated at exactly~\qty{0}{\hertz}.
The parallel plate capacitor has a square side length of~\qty{1}{\milli\metre} and a plate thickness of~\qty{10}{\micro\metre}.
The plate separation was varied between~\qty{40}{\micro\metre} and~\qty{200}{\micro\metre}, and the structure was meshed with~\num{2436} triangles.
The single-turn coil is a ring with a square cross-section of side~\qty{1}{\milli\metre}.
The ring's inner radius was varied from~\qty{10}{\milli\metre} to~\qty{50}{\milli\metre}, with corresponding numbers of mesh triangles ranging from~\num{1626} to~\num{7646}.
To introduce an excitation, a small gap of width~\qty{100}{\micro\metre} was inserted in the ring.
The lumped port excitation described in~\cite{JMMCTFSIE} was used in both cases, which involved appending a row and a column to~\eqref{sys}, representing the coupling to a Th\'evenin-equivalent circuit with a voltage source; details may be found in~\cite{JMMCTFSIE}.

For the parallel plate capacitor, an approximate reference analytical solution was obtained from~\cite[Equation~(11)]{Palmer1937}, which takes fringing fields into account.
For the proposed method, the total charge on each plate was computed by summing the~$\nhat[i]'\cdot\Erp$ unknowns.
The port voltage was computed in post-processing using the approach in~\cite{JMMCTFSIE}, and was found to equal~\qty{1}{\volt} to within~\num{7} decimal places for all plate separations, as expected.
The plate charge and port voltage were then used to find the DC capacitance, shown in \figref{fig_pp} as a function of plate separation for the proposed method, the F-TENH/NETH formulation in~\cite{JMMCTFSIE}, and the reference analytical solution.
Good agreement is observed across all three cases.
The annotation at each datapoint in \figref{fig_pp} shows the number of GMRES iterations required for convergence with the proposed method, confirming numerical stability at DC.

For the ring, an approximate reference analytical solution was used, assuming an infinitesimal ring thickness.
For the proposed method, the magnetic flux was calculated using a simple numerical integration of the magnetic field intercepted by the area enclosed by the ring, ignoring the internal inductance.
The port current was available as a system unknown thanks to the lumped element formulation used~\cite{JMMCTFSIE}.
The magnetic flux and port current together allow direct calculation of the DC 
inductance, shown in the top panel of \figref{fig_ring}, confirming excellent agreement between the proposed method, the formulation in~\cite{JMMCTFSIE}, and the analytical reference.
The resistance was computed using the port voltage and current, and was compared to the simple analytical approximation for a conductor with a rectangular cross section, ignoring the fact that it is bent into a ring.
The lower panel of \figref{fig_ring} shows good agreement across the three methods.
In both panels, the GMRES iteration counts for the proposed method are labelled at each datapoint, confirming DC stability.

\subsection{Sphere Scattering}\label{sec_res_sph}

\begin{figure}[t]
  \centering
  \includegraphics[width=\linewidth]{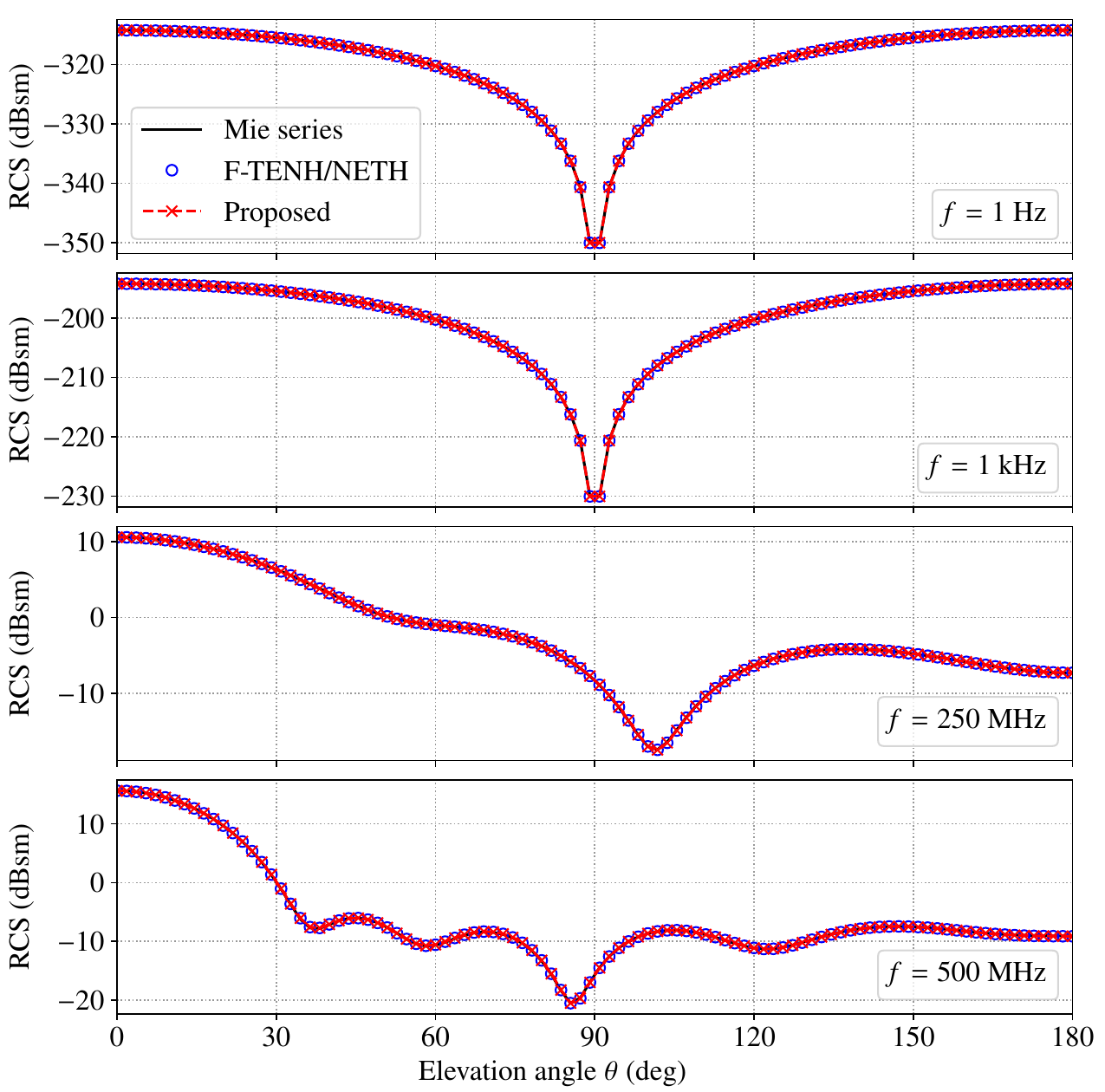}
  \caption{Bistatic radar cross section for the sphere in \secref{sec_res_sph} at different frequencies.}\label{fig_rcssph}
\end{figure}

\begin{figure}[t]
  \centering
  \includegraphics[width=\linewidth]{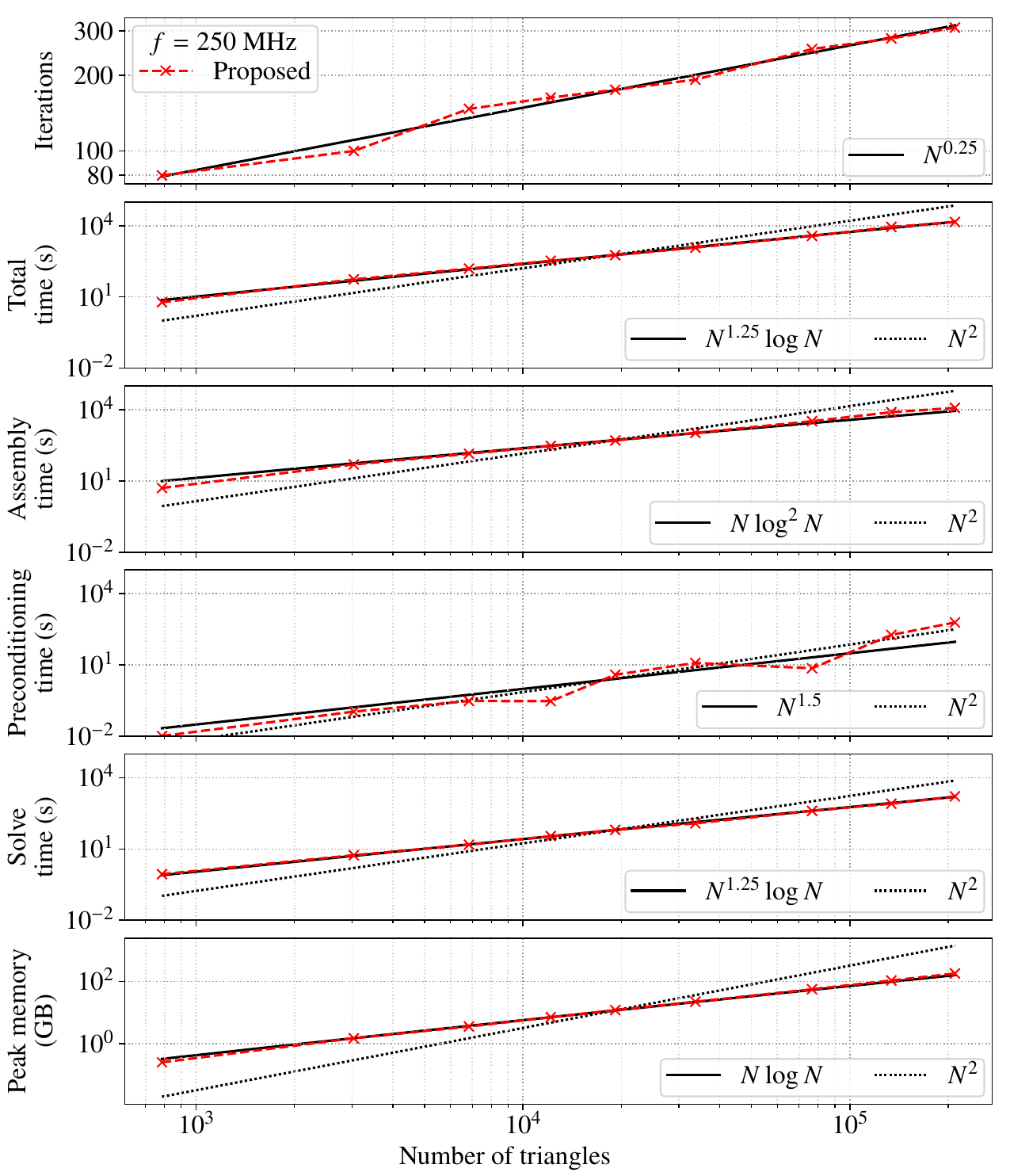}
  \caption{Time, memory, and iteration count scaling with mesh refinement for the sphere in \secref{sec_res_sph} at~\qty{250}{\mega\hertz}.}\label{fig_perfsph}
\end{figure}

The broadband accuracy of the proposed method is verified via scattering from a sphere of diameter~\qty{1}{\meter}, relative permittivity~\num{3}, and electrical conductivity~\qty{0.1}{\siemens\per\meter}, meshed with~\num{6836} triangles.
A plane wave excitation was used, and the bistatic radar cross section was computed as a function of elevation angle in the plane in which the incident electric field was polarized.
\figref{fig_rcssph} confirms good agreement with the analytical Mie series, and with the F-TENH/NETH formulation~\cite{JMMCTFSIE}, from~\qty{1}{\hertz} to the full-wave regime.
For the~\qty{1}{\hertz} case, a GMRES relative residual tolerance of~$10^{-8}$ was used.

To assess the scaling of the proposed method with problem size, the wall-clock time and memory were tracked as a function of mesh refinement, as shown in \figref{fig_perfsph}.
The peak memory usage follows the expected~$\mathcal{O}{(N\log N)}$ trend, where~$N$ is the number of mesh triangles (or equivalently, the number of unknowns).
The top panel of \figref{fig_perfsph} shows the GMRES iteration counts, which grow approximately as~$\mathcal{O}{(N^{0.25})}$ with mesh density.
The ``solve time'' in \figref{fig_perfsph} refers to the iterative solution of~\eqref{sys} with GMRES as described in \secref{sec_accsol}, and follows an~$\mathcal{O}{(N^{1.25}\log N)}$ trend, because the~$\mathcal{O}{(N\log N)}$ cost of each iteration is multiplied by the~$\mathcal{O}{(N^{0.25})}$ growth in iteration counts.
The assembly time, which includes the assembly of all interior and exterior region compressed operators, and the interior-region~\hlu{} procedure described in \secref{sec_acchlu}, is close to~$\mathcal{O}{(N\log^2 N)}$, which is the expected scaling of the interior region~\hlu{} in \secref{sec_acchlu_hlu} up to moderate frequencies.
The solve and assembly time make up the bulk of the total simulation time, which tracks an~$\mathcal{O}{(N^{1.25}\log N)}$ complexity.
The preconditioner assembly and factorization time (referred to as ``preconditioning time'' in \figref{fig_perfsph}) is approximately~$\mathcal{O}{(N^{1.5})}$--$\mathcal{O}{(N^{2})}$, which is within the expectations of sparse factorization algorithms~\cite{superlu,mumps}.
However, even for the largest meshes considered, the preconditioner assembly and factorization take up a small fraction of the total time, so their scaling is unlikely to impact total simulation times even for much larger meshes than the ones considered, although there may eventually be a crossover point where factorization time dominates.

\subsection{Package Interconnects}\label{sec_res_int}

\begin{figure}[t]
  \centering
  \includegraphics[width=\linewidth]{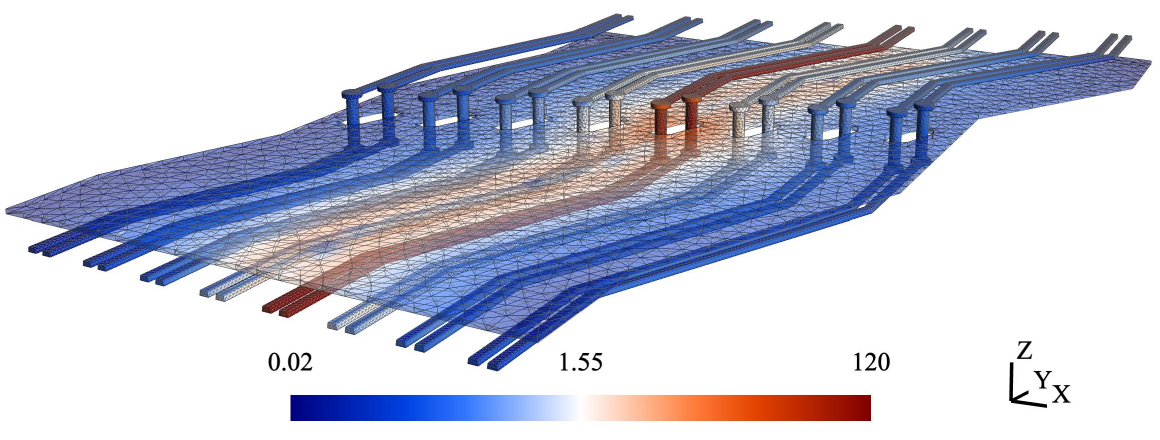}
  \includegraphics[width=\linewidth]{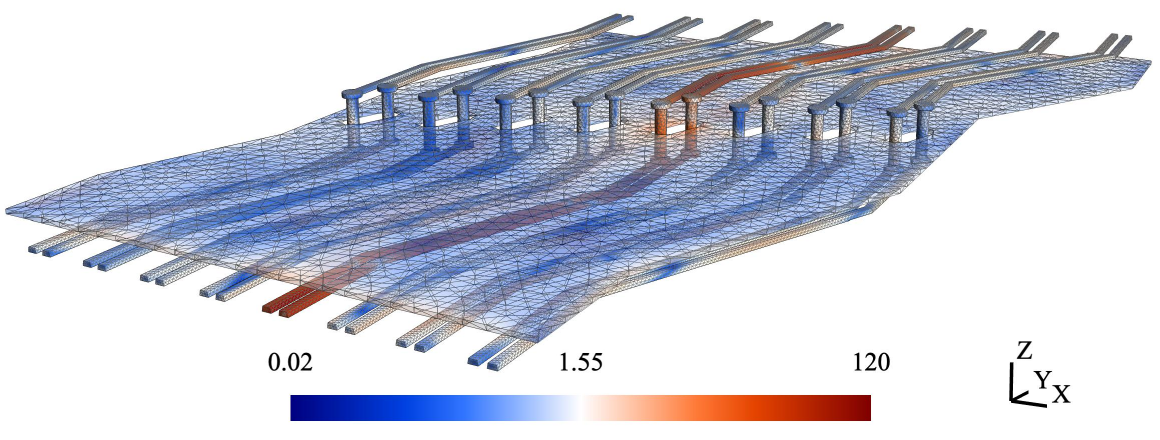}
  \caption{Electric surface current density in~\unit{\ampere\per\meter} for the differential pairs in \secref{sec_res_int} at~\qty{0}{\hertz} (upper panel) and~\qty{100}{\giga\hertz} (lower panel). The ground plane's opacity was reduced to also show the interconnects underneath.}\label{fig_jint}
\end{figure}

\begin{figure}[t]
  \centering
  \includegraphics[width=\linewidth]{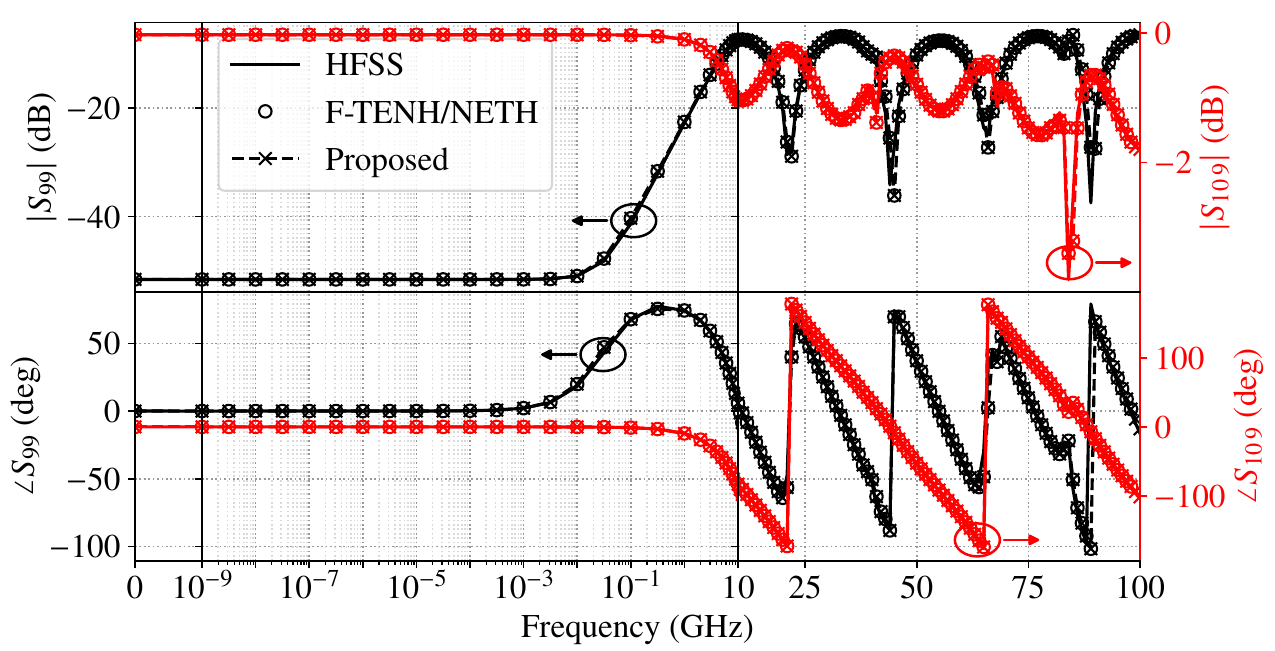}\\
  \includegraphics[width=\linewidth]{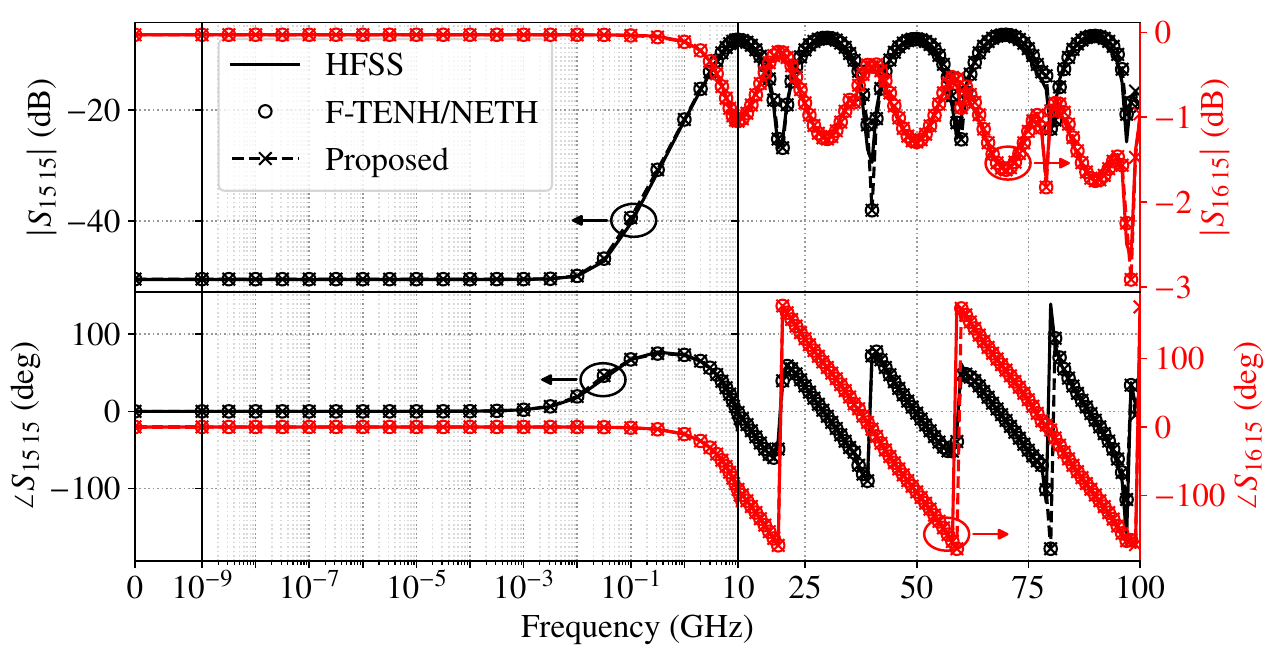}
  \caption{Reflection and transmission scattering parameters for the differential pairs in \secref{sec_res_int}, for a pair near the center of the structure (upper panel), and for the longest pair (lower panel).}\label{fig_sint}
\end{figure}

\begin{figure}[t]
  \centering
  \includegraphics[width=\linewidth]{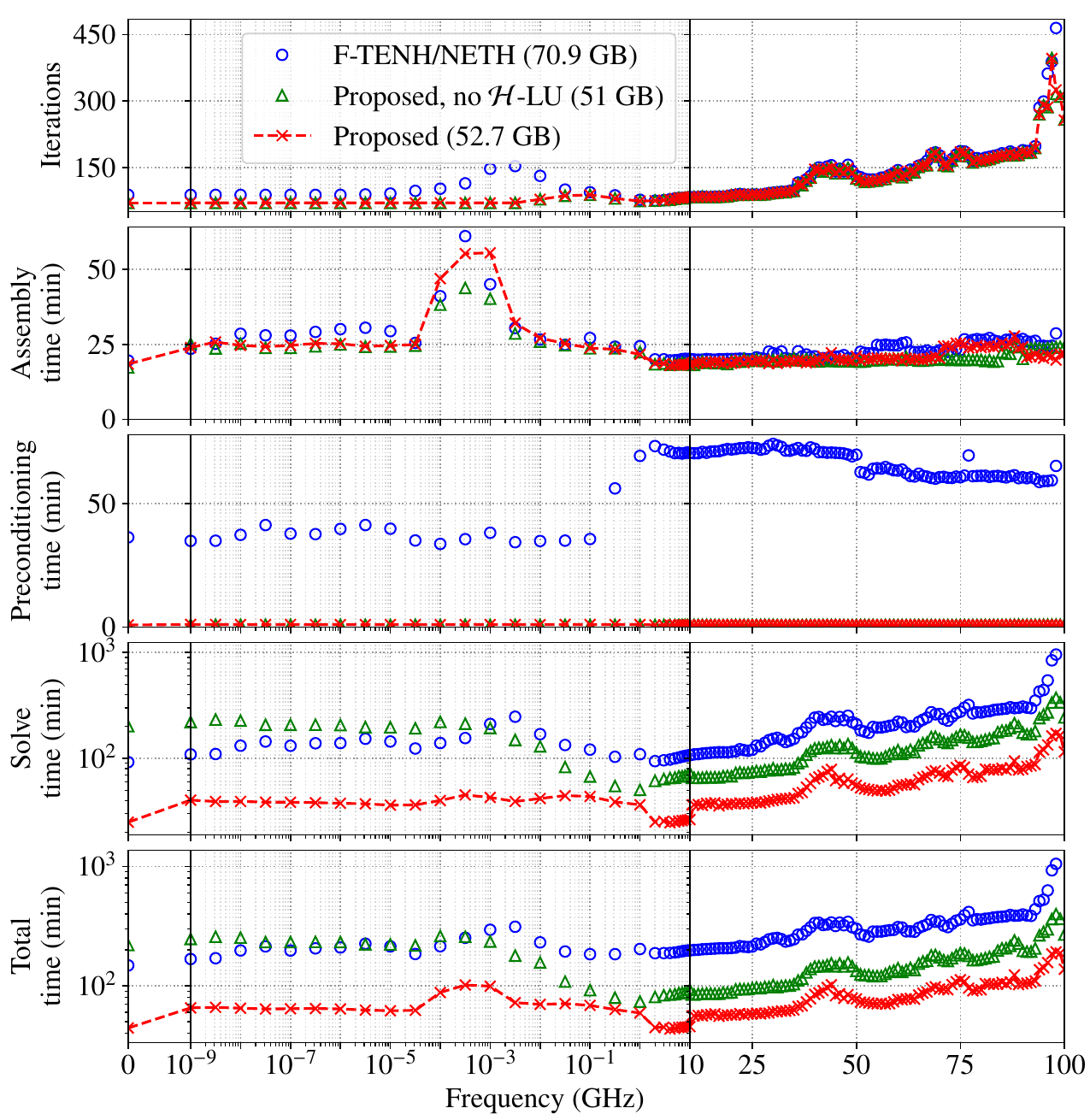}
  \caption{Time, memory, and iteration counts for the differential pairs in \secref{sec_res_int}.}\label{fig_perfint}
\end{figure}

Next, an example representative of high-speed package-level signal integrity analysis is considered, because the simulations in these applications often encompass static, quasistatic, and full-wave regimes in a single frequency sweep.
A metallic structure consisting of eight differential pairs, with cylindrical via transitions through a large common ground plane, is considered.
The metal has a conductivity of~\qty{5.8e7}{\siemens\per\meter}, and the homogeneous background is assumed to have a relative permittivity of~\num{4} to resemble package substrate properties.
Typically, a layered medium Green's function~\cite{BookChew1999} would be used to model the dielectric layers of the package substrate, but a homogeneous background was considered here for simplicity.
Layered medium models can be readily incorporated into the proposed formulation, but are not directly relevant to the goals of this paper, and are highly nontrivial to implement.
The structure was meshed with~\num{72958} triangles.
The proposed method is compared against the F-TENH/NETH formulation~\cite{JMMCTFSIE} and Ansys HFSS~\cite{hfss}.
The lumped port excitation in~\cite{JMMCTFSIE} was used for the proposed method, while de-embedded lumped ports were used in HFSS.

In HFSS, a direct solver was used, with the adaptive meshing tolerance set to~\num{0.01} at~\qty{100}{\giga\hertz}.
To obtain accurate HFSS results at low frequencies down to exactly DC, an interpolating sweep was necessary.
Similar to~\cite{JMMCTFSIE}, and unlike HFSS, the proposed method is applicable as-is at any frequency point, including DC, without invoking low-frequency approximations, interpolation, or solver-switching.
Precluding the need to stitch results from different solver configurations can help avoid passivity and causality violations in the scattering parameters, when analyzing them in the time domain~\cite{Triverio2007}.

The structure and the electric surface current density obtained with the proposed method are visualized in \figref{fig_jint}, at~\qty{0}{\hertz} in the upper panel, and at~\qty{100}{\giga\hertz} in the lower panel.
\figref{fig_sint} shows the reflection and transmission scattering parameters for a pair near the center of the structure, and for the longest pair.
Excellent agreement is observed between all three methods from~\qty{0}{\hertz} to~\qty{100}{\giga\hertz}.
\figref{fig_perfint} shows the simulation time and GMRES iteration counts as a function of frequency for the proposed method, and for the F-TENH/NETH formulation~\cite{JMMCTFSIE}.
The peak memory recorded over the entire sweep is reported in the legend.
Also shown in \figref{fig_perfint} are the performance metrics for the proposed formulation, but using a nested iterative solver for solving~\eqref{sysinx}, similar to the approach in~\cite{TAPAIMin}, instead of the proposed~\hlu{} scheme.
This curve is labeled ``Proposed, no~\hlu{}'' in \figref{fig_perfint}.

The number of GMRES iterations required for the proposed method remains stable and low down to~\qty{0}{\hertz}, but increases at frequencies beyond~\qty{30}{\giga\hertz}, in lock step with the iteration counts for the method in~\cite{JMMCTFSIE}.
The increase in iteration counts with frequency is expected for combined field integral equations with standard RWG-based discretization~\cite{Turc2014,Andriulli2021}.
It has also been observed in augmented formulations~\cite{AWPLSLIM,JMMCTFSIE}, and can be alleviated with advanced preconditioning strategies~\cite{AndriulliConf2018,AndriulliConf2019}, which have not been employed here.
However, high-frequency breakdown explains only the slow increase in iterations with frequency.
The spike in iterations around~\qty{97}{\giga\hertz} in \figref{fig_perfint} seems related to a physical resonance of the ground plane, and disappeared when the ground plane was removed.
It is unlikely to be an \emph{interior} resonance, because the proposed method makes use of combined field formulations, which are a well-known remedy~\cite{BookChew2009}.
A possible explanation might be the proposed preconditioner in \secref{sec_pc}, which is based on those in~\cite{Chew2009,JMMCTFSIE}.
The preconditioner retains self-interactions of each square integral operator, which are expected to be dominant.
However, in a finite-thickness, thin, large ground plane, interactions between the bottom and top faces are likely to be both numerous and strong.
The proposed preconditioner takes into account the connectivity of the mesh, but not the proximity between unconnected elements.
Not capturing these strong nearby interactions in the preconditioner may have a large impact on iteration counts at physical resonances of the structure.
A distance-based preconditioning strategy might be a better choice, but would substantially increase the computational cost.
Since the overall impact on total simulation time is negligible, as seen in \figref{fig_perfint}, and the main claims of the article remain unchanged, this topic is deferred to future work.

The assembly time in \figref{fig_perfint} is comparable across all three methods because they build the same set of compressed operators.
For the proposed method, this includes the~\hlu{} step from \secref{sec_acchlu_hlu}, but the time associated with the~\hlu{} factorizations is small compared to the assembly of the~\hmats{} themselves, because the~\hlu{} is performed per object rather than across the entire structure.
The sparse preconditioner in \secref{sec_pc} is substantially quicker to factorize than the one for the method in~\cite{JMMCTFSIE}, because the former is half the size, and has fewer than a quarter of the number of non-zeros by construction, which also has significant downstream impacts on fill-ins.
Across all frequencies, preconditioner assembly and factorization with the proposed method is~\num{33}$\times$ to~\num{90}$\times$ faster than the one in~\cite{JMMCTFSIE}.
Considering the discussion around preconditioner time scaling in \secref{sec_res_sph}, the proposed method is expected to be able to handle much larger problems than the formulation in~\cite{JMMCTFSIE} before becoming bottlenecked by the preconditioner.

The solve time in \figref{fig_perfint} highlights the benefits of the proposed~\hlu{} scheme for computing the immittance operators, compared to using a nested iterative solver as in~\cite{TAPAIMin}.
The latter requires running an inner GMRES procedure for each object, at each outer GMRES iteration, which is not only slower, but also makes the total solve time more susceptible to the condition number of the interior region system in~\eqref{sysinx}.
Compared to~\cite{JMMCTFSIE}, across all frequencies, the solve stage with the proposed method is~\num{2.7}$\times$ to~\num{6.3}$\times$ faster.

Overall, the proposed method is~\num{2.4}$\times$ to~\num{5.5}$\times$ faster than the F-TENH/NETH formulation~\cite{JMMCTFSIE} across the sweep, as shown in the bottom panel of \figref{fig_perfint}.
At the same time, the proposed method remains accurate from exactly~\qty{0}{\hertz} to millimeter-wave frequencies.
However, since these savings come from having a smaller system size, the proposed method does not necessarily have a \emph{scaling} advantage compared to~\cite{JMMCTFSIE}.

\subsection{Metastructure-Inspired Array}\label{sec_res_arr}
 
\begin{figure}[t]
  \centering
  \includegraphics[width=.45\linewidth]{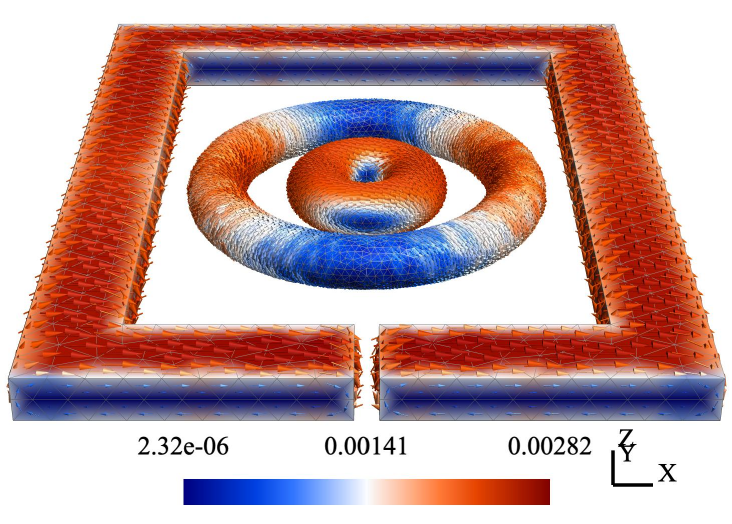}
  \includegraphics[width=.45\linewidth]{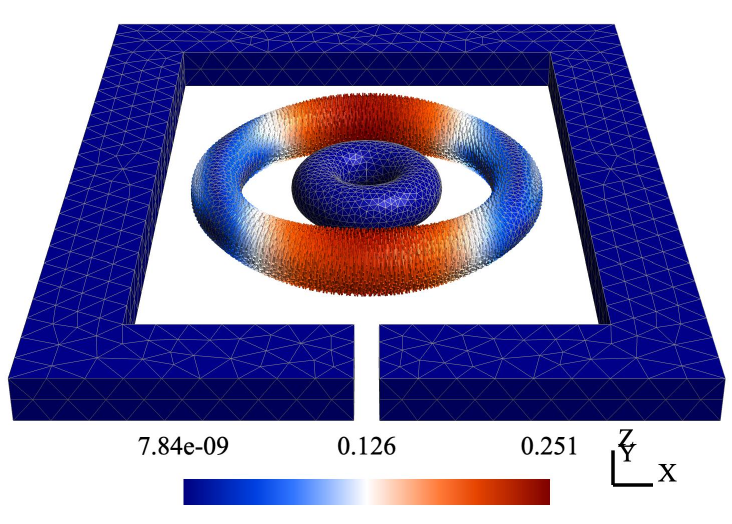}
  \caption{Electric (left,~\unit{\ampere\per\meter}) and magnetic (right,~\unit{\volt\per\meter}) surface current density for the unit cell in \secref{sec_res_arr} at~\qty{0}{\hertz}.}\label{fig_jmsrr0}
\end{figure}

\begin{figure}[t]
  \centering
  \includegraphics[width=\linewidth]{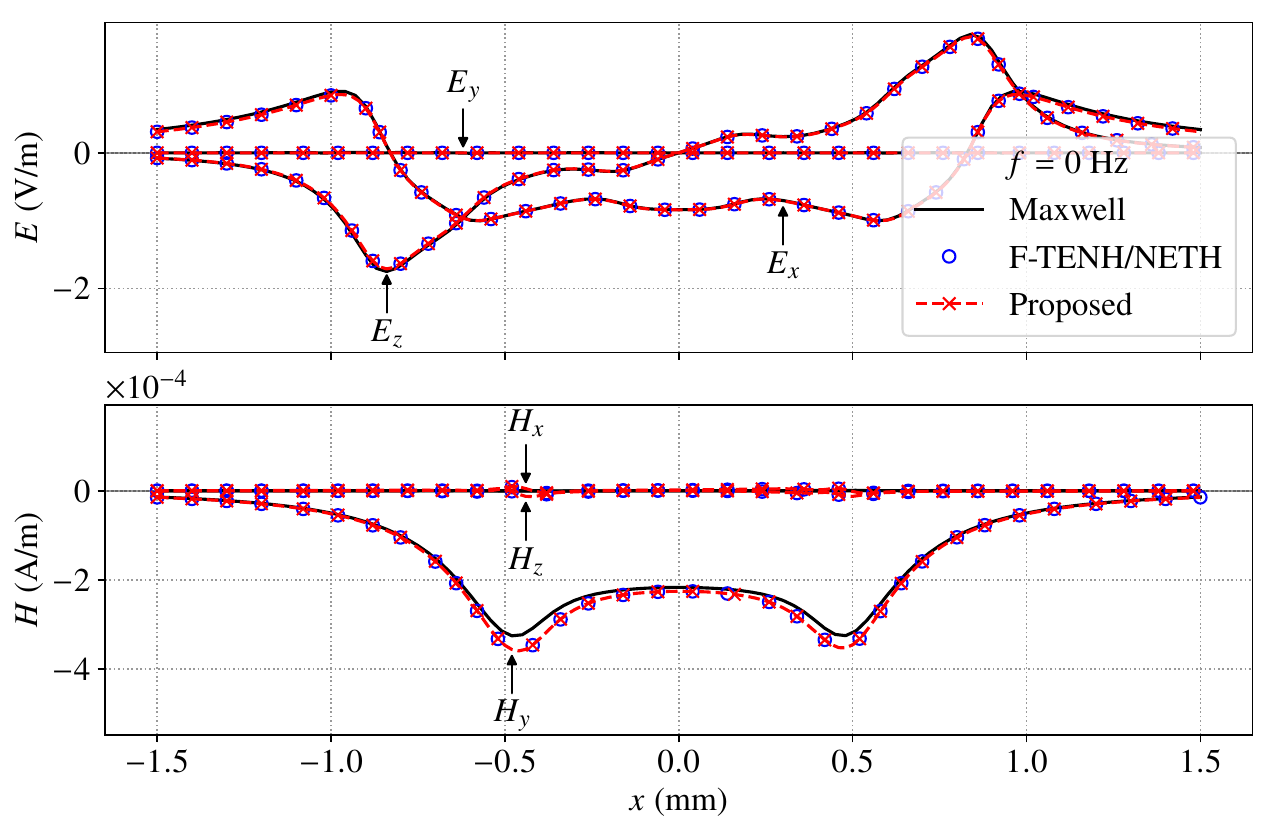}
  \caption{Electric and magnetic near fields for the unit cell in \secref{sec_res_arr} at~\qty{0}{\hertz}.}\label{fig_esrr0}
\end{figure}

\begin{figure}[t]
  \centering
  \includegraphics[width=\linewidth]{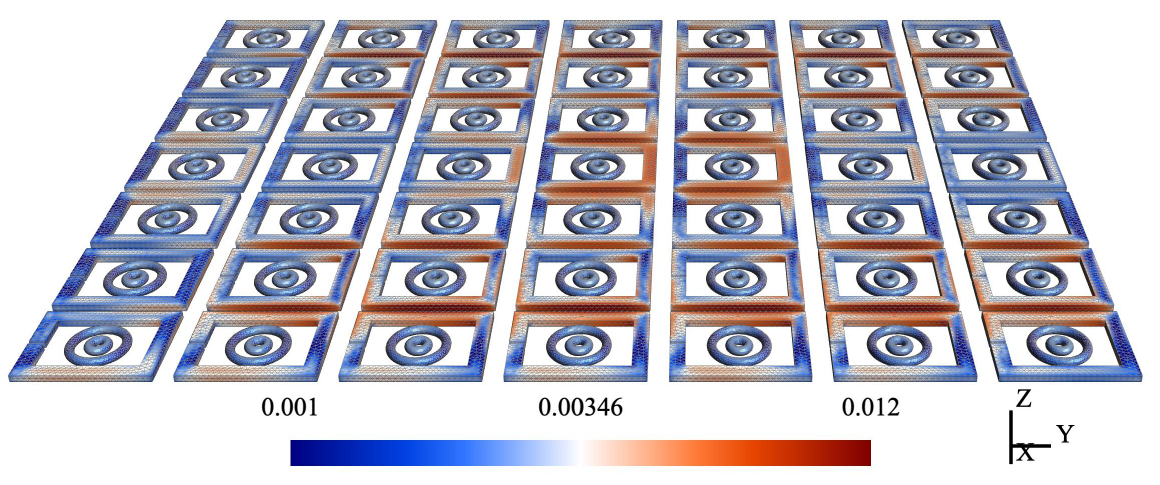}
  \caption{Electric surface current density in~\unit{\ampere\per\meter} for the~\numproduct{7x7} array in \secref{sec_res_arr} at~\qty{30}{\giga\hertz}, under plane wave incidence.}\label{fig_jsrr}
\end{figure}

\begin{figure}[t]
  \centering
  \includegraphics[width=\linewidth]{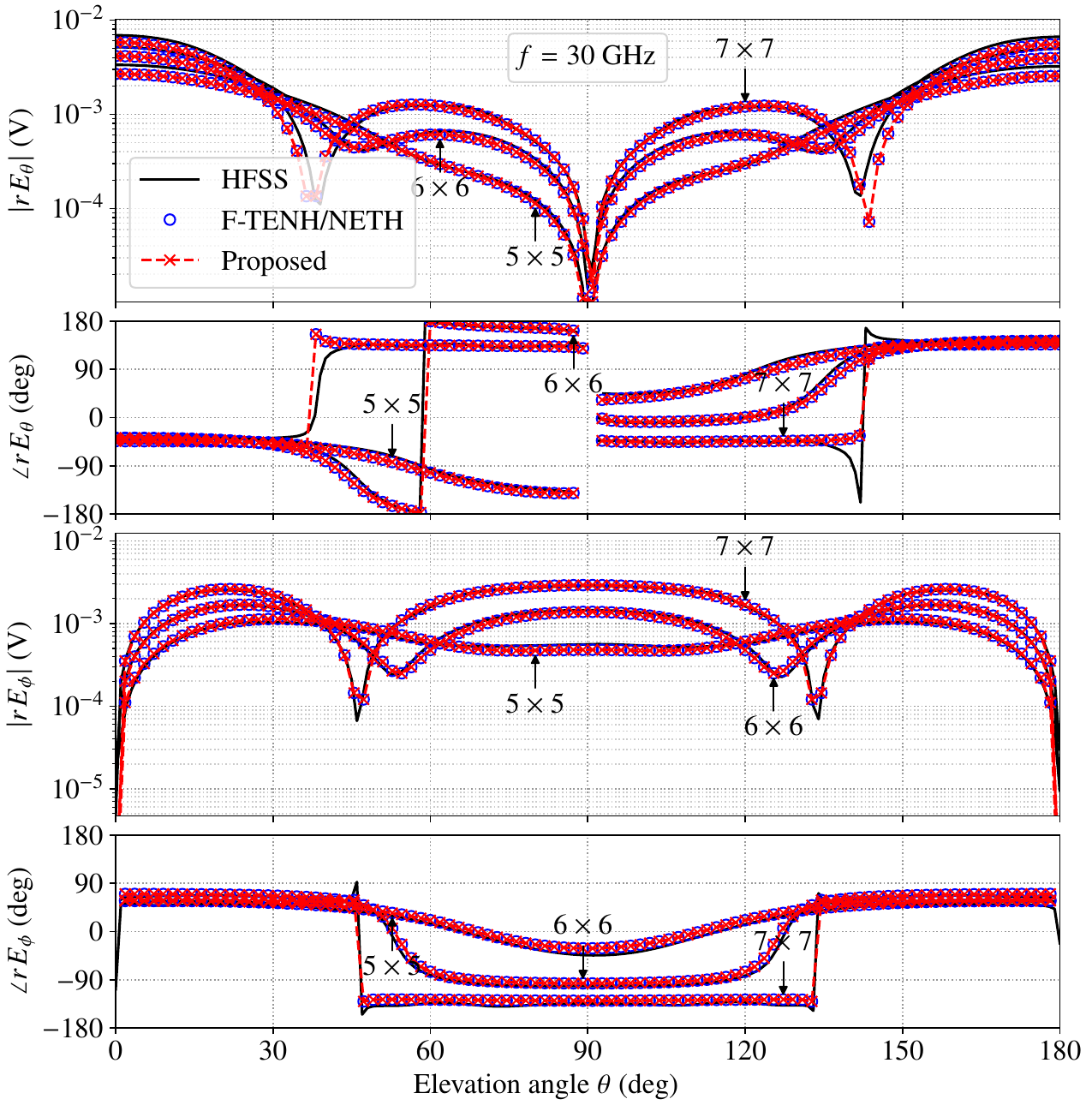}
  \caption{Electric far field components for the arrays in \secref{sec_res_arr} at~\qty{30}{\giga\hertz}, where~$r$ is the radial measurement distance from the structure's centroid.}\label{fig_esrr}
\end{figure}

\begin{figure}[t]
  \centering
  \includegraphics[width=\linewidth]{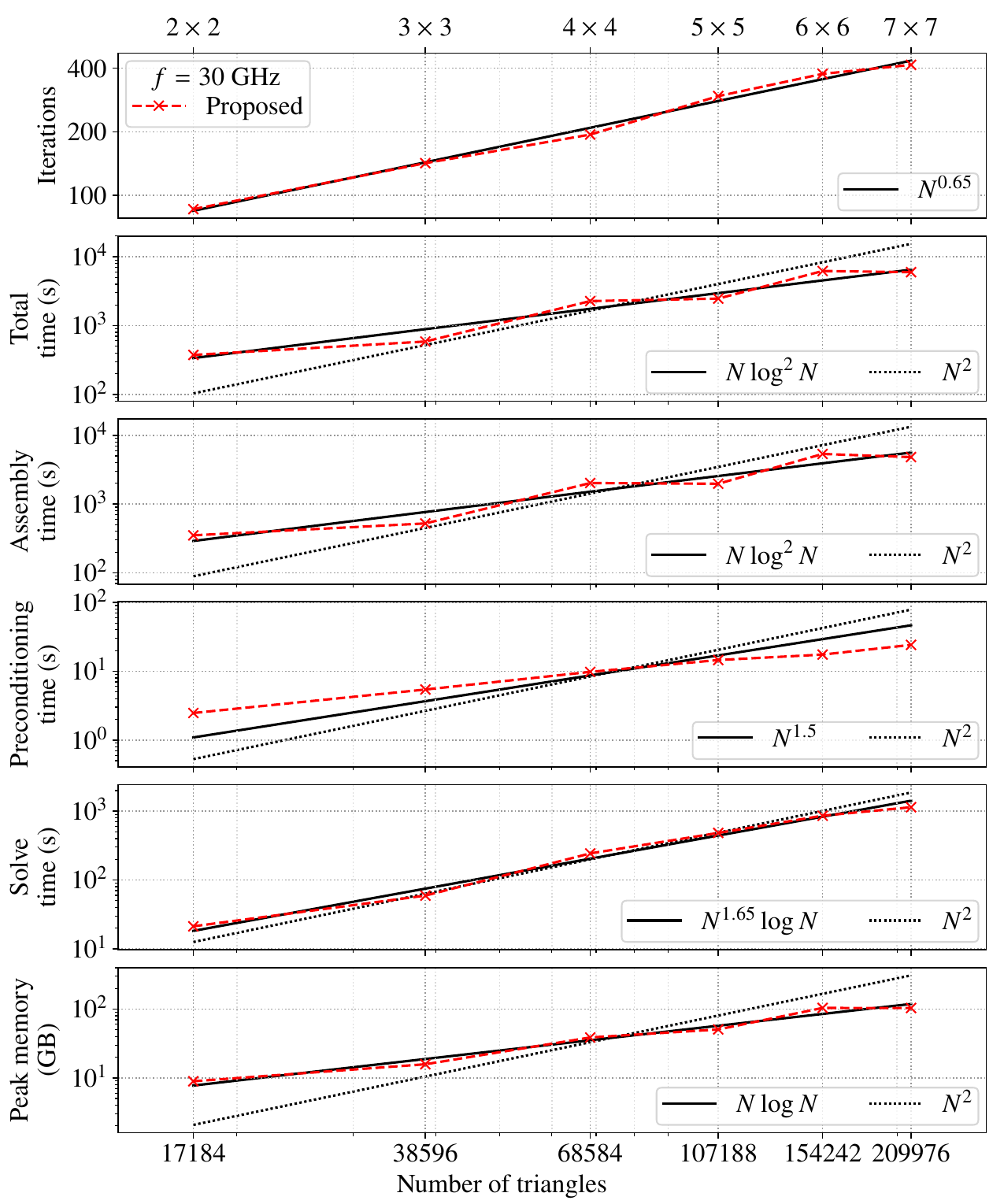}
  \caption{Time, memory, and iteration count scaling with array size for the arrays in \secref{sec_res_arr} at~\qty{30}{\giga\hertz}.}\label{fig_perfsrr}
\end{figure}

An array of objects representative of metamaterial and metasurface unit cells is considered next.
Each unit cell consists of a torus with relative permittivity~\num{12} and conductivity~\qty{10}{\siemens\per\meter}, encircled by a torus of relative permittivity~\num{4} and relative permeability~\num{10}, which lies inside a rectangular split ring resonator with conductivity~\qty{e4}{\siemens\per\meter}.

First, the accuracy of the proposed method at~\qty{0}{\hertz} is verified for a single unit cell under plane wave illumination.
At~\qty{0}{\hertz}, the plane wave is simply a superposition of uniform static electric and magnetic fields.
The electric and magnetic surface current densities for the proposed method are shown in \figref{fig_jmsrr0}, when the impressed electric field is polarized along the~$x$ axis.
The scattered near field is measured along a line parallel to the~$x$ axis,~\qty{75}{\micro\meter} above the structure, passing directly over the common center of the two tori.
As validation, Ansys Maxwell~\cite{maxwell} was used, in which separate electrostatic and magnetostatic simulations were required, with appropriately set up field directions and amplitudes to match the plane wave used in the proposed method.
The electric and magnetic near fields are reported in~\figref{fig_esrr0}, showing good agreement between the proposed method, the F-TENH/NETH formulation~\cite{JMMCTFSIE}, and the two Maxwell simulations.

Next, the scattered far fields and performance are verified by considering arrays ranging from~\numproduct{2x2} to~\numproduct{7x7} unit cells.
The~\numproduct{7x7} case with its electric surface current density at~\qty{30}{\giga\hertz} obtained with the proposed method is visualized in \figref{fig_jsrr}.
The excitation was a plane wave traveling along the~$z$ axis, with the electric field polarized along the~$x$ axis.
\figref{fig_esrr} shows the electric far field at~\qty{30}{\giga\hertz} for the~\numproduct{5x5},~\numproduct{6x6}, and~\numproduct{7x7} cases in~$xz$ plane.
The upper two panels show the magnitude and phase, respectively, for the elevation component~$E_\theta$, while the lower two panels show the azimuthal component~$E_\phi$, and good agreement with HFSS is observed.

\figref{fig_perfsrr} shows the time, memory, and iteration count scaling of the proposed method as a function of array size.
In contrast with the study in \secref{sec_res_sph}, here the electrical size of the structure is varied for a fixed mesh density.
Overall, the scaling is similar to that observed in \secref{sec_res_sph}.
Preconditioning time scales better in this example than in \secref{sec_res_sph}, likely because the structure has many small components rather than a single large object, which leads to an overall sparser preconditioner matrix, thus faster factorization.
The solve time follows an~$\mathcal{O}{(N^{1.65}\log N)}$ trend in this case, because the iteration count grows approximately as~$\mathcal{O}{(N^{0.65})}$ with array size, compared to~$\mathcal{O}{(N^{0.25})}$ in \secref{sec_res_sph}.
The total time follows an~$\mathcal{O}{(N\log^2 N)}$ trend because it is dominated by the assembly time, but it will approach the scaling of the solve time if the number of excitations is increased such that solve time dominates.
For cases with an even number of unit cells per side, the exterior-region~\hmat{} clusters divide the unit cells perfectly evenly among them, which causes the extended admissibility criterion~\cite{Mao2018} to create a few large admissible blocks compared to the odd-numbered cases.
This causes the assembly time fluctuations observed in \figref{fig_perfsrr}, where the time for odd-numbered arrays can be comparable to the time for the next-smallest even-numbered array.
Compared to the F-TENH/NETH formulation~\cite{JMMCTFSIE} across all array sizes, the proposed method had a similar assembly time, was~{\num{21}$\times$--\num{32}$\times$} faster in preconditioner assembly and factorization, and~{\num{2.1}$\times$--\num{5.1}$\times$} faster in solve time for a single excitation.
Since the assembly time dominates in this case, the proposed formulation is only~{\num{1.2}$\times$--\num{1.6}$\times$} faster in total time, but this ratio increases in favor of the proposed approach as the number of excitations increases and solve time begins to dominate.
For example, with~\num{10} excitations, the proposed method is up to~{\num{3.2}$\times$} faster across the cases.

\section{Conclusion}\label{sec_concl}

A new immittance operator formulation was proposed, which enables the unified electromagnetic simulation of dielectrics, semiconductors, and lossy metals from exactly~\qty{0}{\hertz} to the full-wave regime.
A matrix compression strategy for the proposed approach was devised, which leverages hierarchical ($\mathcal{H}$)-matrices to provide an overall nearly log-linear time and memory scaling.
The proposed method outperforms the existing frequency-stable F-TENH/NETH formulation~\cite{JMMCTFSIE}, but retains its broadband numerical stability and accuracy.
Numerical examples drawn from different applications demonstrated the accuracy and efficiency of the proposed method, including its nearly log-linear time and memory complexity.

\section{Acknowledgment}\label{sec_ack}

Some of the numerical tests in \secref{sec_res} were executed on the University of Birmingham's BlueBEAR high-performance computing service.

Artificial intelligence (Claude Opus) was used to:
\begin{itemize}
    \item check for typos and math inconsistencies such as missing subscripts and notation mistakes;
    \item check for missing data or typographical mistakes in bibliography fields;
    \item suggest rewording of a small number of sentences (fewer than~\qty{1}{\percent} of the total), based on text written by the author; all such text was reviewed and revised by the author;
    \item help write Python plotting scripts;
    \item help write Python scripts to create the Gmsh geometries in \secref{sec_res_int} and \secref{sec_res_arr};
    \item help write an in-house~\hlu{} implementation and validate it against HLIBpro~\cite{hlibpro}.
\end{itemize}

{\appendices
\section*{Appendix}
The matrix operators in~\eqref{sys} are defined as follows, for the~$l^{\mathrm{th}}$ testing function and~$m^{\mathrm{th}}$ expansion function, where~$\vec{f}_l(\vec{r})$ denotes the edge-length-normalized RWG function associated with mesh edge~$l$, and~$h_l(\vec{r})$ denotes the area-normalized piecewise constant function associated with mesh triangle~$l$, both on~$\Gamma_i$:
\begin{align}
  \matr{L}_{i,(l,m)}^{(\mathrm{t})} &= \surfint[i]\,\vec{f}_l(\vec{r})\cdot\surfintp[i]\,\Grrpki[i]\,\vec{f}_m(\vec{r}\,'),\label{opLt}
\end{align}
\begin{align}
  \matr{L}_{i,(l,m)}^{(\mathrm{s})} &= \surfint[i]\,h_l(\vec{r})\surfintp[i]\,\Grrpki[i]\,h_m(\vec{r}\,'),\label{opLs}
\end{align}
\begin{align}
  \matr{L}_{i,(l,m)}^{(\mathrm{r})} &= \surfint[i]\,(\nhat[i]\times\vec{f}_l(\vec{r}))\cdot\surfintp[i]\,\Grrpki[i]\,\vec{f}_m(\vec{r}\,'),\label{opLr}
\end{align}
\begin{align}
  \widetilde{\matr{K}}_{i}^{(\mathrm{t})} &= \frac{1}{2}\matr{I}_{i}^{(\mathrm{r})} + \matr{K}_{i}^{(\mathrm{t})},\label{optKt}\\
  \widetilde{\matr{K}}_{i}^{(\mathrm{r})} &= \frac{1}{2}\matr{I}_{i}^{(\mathrm{t})} - \matr{K}_{i}^{(\mathrm{r})},\label{optKr}
\end{align}
\begin{align}
  \matr{K}_{i,(l,m)}^{(\mathrm{t})} &= \surfint[i]\,\vec{f}_l(\vec{r})\cdot\pvsurfintp[i]\,\nabla\Grrpki[i]\times\vec{f}_m(\vec{r}\,'),\label{opKt}
\end{align}
\begin{align}
  \begin{split}
    \matr{K}_{i,(l,m)}^{(\mathrm{r})} &= \surfint[i]\,(\nhat[i]\times\vec{f}_l(\vec{r}))\cdot\\ &\hspace{50pt}\pvsurfintp[i]\,\nabla\Grrpki[i]\times\vec{f}_m(\vec{r}\,'),\label{opKr}
  \end{split}
\end{align}
\begin{align}
  \matr{I}_{i,(l,m)}^{(\mathrm{r})} &= \surfint[i]\,(\nhat[i]\times\vec{f}_l(\vec{r}))\cdot\vec{f}_m(\vec{r}),\label{opIr}\\
  \matr{I}_{i,(l,m)}^{(\mathrm{t})} &= \surfint[i]\,\vec{f}_l(\vec{r})\cdot\vec{f}_m(\vec{r}),\label{opIt}
\end{align}
\begin{align}
  \begin{split}
    \matr{M}_{i,(l,m)} &= \surfint[i]\,(\nhat[i]\times\vec{f}_l(\vec{r}))\cdot\\
                       &\hspace{50pt}\pvsurfintp[i]\,\nabla\Grrpki[i]\,h_m(\vec{r}\,'),\label{opM}
  \end{split}\\
  \matr{D}_{i,(l,m)} &=
                       \begin{cases}
                         1 & \text{$m$ positively polarized w.r.t. $l$},\\
                         -1 & \text{$m$ negatively polarized w.r.t. $l$},\\
                         0 & \text{otherwise},
                       \end{cases}\label{opD}
\end{align}
where
${\matr{L}_{i}^{(\mathrm{t})}, \matr{K}_{i}^{(\mathrm{t})}, \matr{L}_{i}^{(\mathrm{r})}, \matr{K}_{i}^{(\mathrm{r})}, \matr{I}_{i}^{(\mathrm{t})}, \matr{I}_{i}^{(\mathrm{r})}  \in \mathbb{C}^{N_{\mathrm{edge},i}\times N_{\mathrm{edge},i}}}$,
${\matr{L}_{i}^{(\mathrm{s})} \in \mathbb{C}^{N_{\mathrm{face},i}\times N_{\mathrm{face},i}}}$,
${\matr{M}_{i} \in \mathbb{C}^{N_{\mathrm{edge},i}\times N_{\mathrm{face},i}}}$,
and ${\matr{D}_{i} \in \mathbb{Z}^{N_{\mathrm{face},i}\times N_{\mathrm{edge},i}}}$.
Integers~$N_{\mathrm{edge},i}$ and~$N_{\mathrm{face},i}$ denote the number of mesh edges and faces (triangles), respectively.
In~\eqref{opKt},~\eqref{opKr}, and~\eqref{opM}, the symbol~$\pvint$ denotes an integral in the principal value sense.

The right-hand side vectors in~\eqref{sysout} and~\eqref{sys} are the tested right-hand sides of~\eqref{tefie} and~\eqref{nmfie} for~${i=0}$,
\begin{align}
    \matr{b}_{l}^{(\mathrm{t})} &= \surfint[0]\,\vec{f}_l(\vec{r})\cdot\vec{E}^{\mathrm{inc}}(\vec{r}),\label{rhsbt}\\
    \matr{b}_{l}^{(\mathrm{r})} &= \surfint[0]\,\vec{f}_l(\vec{r})\cdot\left(\nhat[0]\times\vec{H}^{\mathrm{inc}}(\vec{r})\right),\label{rhsbr}
\end{align}
where
\begin{align}
    \vec{E}^{\mathrm{inc}}(\vec{r}) &= -j\omega\mu_0\,\mathcal{V}_0\left[\Jrp[\mathrm{im}]\right](\vec{r}) - \frac{1}{\eps_0}\,\nabla\mathcal{V}_0\left[\rhorp[\mathrm{im}]\right](\vec{r}),\label{einc}\\
    \vec{H}^{\mathrm{inc}}(\vec{r}) &= \curl\mathcal{V}_0\left[\Jrp[\mathrm{im}]\right](\vec{r})\label{hinc}
\end{align}
are the fields radiated in the homogeneous background by the impressed sources in~$\Omega_0$, in the absence of the structure.
For a plane wave of amplitude~$E_0$, polarization~$\hat{e}$, and propagation direction~$\hat{k}$,
\begin{align}
  \vec{E}^{\mathrm{inc}}(\vec{r}) &= E_0\,\hat{e}\,e^{-jk_0\,\hat{k}\cdot\vec{r}},\label{pwe}\\
  \vec{H}^{\mathrm{inc}}(\vec{r}) &= \frac{1}{\eta_0}\,\hat{k}\times\vec{E}^{\mathrm{inc}}(\vec{r}),\label{pwh}
\end{align}
where~${\eta_0=\sqrt{\mu_0/\eps_0}}$ is the background wave impedance.
For lumped port excitations, both vectors vanish, and the excitation enters instead through the port equations appended to~\eqref{sys}~\cite{JMMCTFSIE}.
Here,~${\matr{b}^{(\mathrm{t})}, \matr{b}^{(\mathrm{r})} \in \mathbb{C}^{N_{\mathrm{edge},0}}}$.
}

\bibliographystyle{IEEEtran}
\bibliography{IEEEabrv,bibliography}

@STRING{IEEE_J_ADVP       = "{IEEE} Trans. Adv. Packag."}

@STRING{IEEE_J_AWPL       = "{IEEE} Antennas Wireless Propag. Lett."}

@STRING{IEEE_J_AP         = "{IEEE} Trans. Antennas Propag."}

@STRING{IEEE_J_EMC        = "{IEEE} Trans. Electromagn. Compat."}

@STRING{IEEE_J_MTT        = "{IEEE} Trans. Microw. Theory Tech."}

@STRING{IEEE_J_MMCT       = "{IEEE} J. Multiscale Multiphys. Comput. Tech."}

@STRING{IEEE_M_AP         = "{IEEE} Antennas Propag. Mag."}

@STRING{IEEE_M_MW         = "{IEEE} Microw. Mag."}

@book{BookChew2009,
    title = {Integral Equation Methods for Electromagnetic and Elastic Waves},
    publisher = {Springer},
    address = {New York, NY},
    year = {2009},
    author = {Weng Cho Chew and Mei Song Tong and Bin Hu},
    edition = {First}
}

@book{BookChew1999,
    title = {Waves and Fields in Inhomogeneous Media},
    publisher = {Wiley-IEEE Press},
    address = {Piscataway, NJ},
    year = {1999},
    author = {Weng Cho Chew}
}

@book{BookBode1945,
    title = {Network Analysis and Feedback Amplifier Design},
    publisher = {Van Nostrand},
    address = {New York, NY},
    year = {1945},
    author = {Hendrik W. Bode}
}

@book{BookHanson2002,
    title = {Operator Theory for Electromagnetics},
    publisher = {Springer},
    address = {New York, NY},
    year = {2002},
    author = {George W. Hanson and Alexander B. Yakovlev}
}

@book{BookGurel2014,
    title = {The Multilevel Fast Multipole Algorithm ({MLFMA}) for Solving Large-Scale Computational Electromagnetics Problems},
    publisher = {Wiley-IEEE Press},
    address = {Piscataway, NJ},
    year = {2014},
    author = {{\"O}zg{\"u}r Erg{\"u}l and Levent G{\"u}rel}
}

@book{BookHackbusch2014,
    title = {Hierarchical Matrices: Algorithms and Analysis},
    publisher = {Springer},
    address = {Berlin},
    year = {2015},
    author = {Wolfgang Hackbusch},
}

@book{BookYuferevIda2010,
    title = {Surface Impedance Boundary Conditions: A Comprehensive Approach},
    publisher = {CRC Press},
    address = {Boca Raton, Florida},
    year = {2010},
    author = {Sergey V. Yuferev and Nathan Ida}
}

@article{superlu,
    author = {Li, Xiaoye S.},
    title = {An overview of {SuperLU}: Algorithms, implementation, and user interface},
    year = {2005},
    month = {Sep.},
    publisher = {Assoc. Comput. Mach.},
    address = {New York, NY, USA},
    volume = {31},
    number = {3},
    doi = {10.1145/1089014.1089017},
    journal = {ACM Trans. Math. Softw.},
    pages = {302-325}
}

@InProceedings{mumps,
    author={Amestoy, Patrick R. and Duff, Iain S. and L'Excellent, Jean-Yves and Koster, Jacko},
    editor={S{\o}revik, Tor and Manne, Fredrik and Gebremedhin, Assefaw Hadish and Moe, Randi},
    title={{MUMPS}: A General Purpose Distributed Memory Sparse Solver},
    booktitle={Applied Parallel Computing. New Paradigms for {HPC} in Industry and Academia},
    year={2001},
    publisher={Springer},
    address={Berlin, Heidelberg},
    pages={121-130},
}

@article{gmres,
    author = {Saad, Youcef and Schultz, Martin H},
    title = {{GMRES}: a generalized minimal residual algorithm for solving nonsymmetric linear systems},
    year = {1986},
    volume = {7},
    number = {3},
    journal = {SIAM J. Sci. Stat. Comput.},
    month = {Jul.},
    pages = {856-869},
}

@misc{hlibpro,
    author = {Kriemann, R.},
    title = {{HLIBpro}},
    howpublished = {\url{https://www.hlibpro.com}},
    note = {Accessed: Nov. 3, 2026}
}

@ARTICLE{Palmer1937,
    author={Palmer, Harlan B.},
    journal={Elect. Eng.}, 
    title={The capacitance of a parallel-plate capacitor by the {Schwartz-Christoffel} transformation}, 
    month={Mar.},
    year={1937},
    volume={56},
    number={3},
    pages={363-368},
    doi={10.1109/EE.1937.6540485}
}

@ARTICLE{Matthaei1954,
    author={Matthaei, George L.},
    journal={Proc. IRE}, 
    title={Some Techniques for Network Synthesis}, 
    month={Jul.},
    year={1954},
    volume={42},
    number={7},
    pages={1126-1137},
}

@ARTICLE{Triverio2007,
    author={Triverio, Piero and Grivet-Talocia, Stefano and Nakhla, Michel S. and Canavero, Flavio G. and Achar, Ramachandra},
    journal=IEEE_J_ADVP,
    title={Stability, Causality, and Passivity in Electrical Interconnect Models},
    year={2007},
    volume={30},
    number={4},
    month={Nov.},
    pages={795-808},
    doi={10.1109/TADVP.2007.901567}
}

@article{StrattonChu1939,
    title = {Diffraction Theory of Electromagnetic Waves},
    author = {Stratton, J. A. and Chu, L. J.},
    journal = {Phys. Rev.},
    volume = {56},
    number = {1},
    pages = {99-107},
    numpages = {0},
    year = {1939},
    month = {Jul.},
    publisher = {American Physical Society},
    doi = {10.1103/PhysRev.56.99},
}

@article{RWG1982,
    author={Rao, S. and Wilton, D. and Glisson, A.},
    journal=IEEE_J_AP,
    title={Electromagnetic scattering by surfaces of arbitrary shape},
    year={1982},
    volume={30},
    month={May},
    number={3},
    pages={409-418},
    doi={10.1109/TAP.1982.1142818}
}

@article{Vecchi1999,
    author={Vecchi, G.},
    journal=IEEE_J_AP,
    title={Loop-star decomposition of basis functions in the discretization of the {EFIE}},
    year={1999},
    volume={47},
    number={2},
    pages={339-346},
    month={Feb.},
    doi={10.1109/8.761074}
}

@article{Bebendorf2000,
    author = {Bebendorf, M.},
    doi = {10.1007/PL00005410},
    journal = {Numer. Math.},
    number = {4},
    pages = {565--589},
    title = {Approximation of Boundary Element Matrices},
    volume = {86},
    month = {Oct.},
    year = {2000},
}

@article{Grasedyck2005,
    author = {Grasedyck, L.},
    journal = {Computing},
    month = {May},
    number = {3},
    pages = {205--223},
    title = {Adaptive Recompression of {$\mathcal{H}$-}Matrices for {BEM}},
    volume = {74},
    year = {2005},
}

@article{BebendorfRjasanow2003,
    author = {Bebendorf, M. and Rjasanow, S.},
    journal = {Computing},
    month = {Feb.},
    number = {1},
    pages = {1--24},
    title = {Adaptive Low-Rank Approximation of Collocation Matrices},
    volume = {70},
    year = {2003},
}

@article{BebendorfKunis2009,
    author = {M. Bebendorf and S. Kunis},
    journal = {J. Integral Equ. Appl.},
    number = {3},
    pages = {331--357},
    title = {Recompression Techniques for Adaptive Cross Approximation},
    volume = {21},
    year = {2009}
}

@article{JFLee2003,
    author={Jin-Fa Lee and Lee, R. and Burkholder, R. J.},
    journal=IEEE_J_AP,
    title={Loop star basis functions and a robust preconditioner for {EFIE} scattering problems},
    year={2003},
    volume={51},
    number={8},
    month={Aug.},
    pages={1855-1863},
    doi={10.1109/TAP.2003.814736}
}

@ARTICLE{Ren2024,
    author={Ren, Yi and Chen, Yongpin and Liu, Yanhui and Wang, Jinhong},
    journal=IEEE_J_AWPL, 
    title={Fast Direct Solution of Composite Dielectric Objects by {M-HODLR}}, 
    year={2024},
    month={Jan.},
    volume={23},
    number={1},
    pages={254-258},
    doi={10.1109/LAWP.2023.3322737}
}

@ARTICLE{Xie2022,
    author={Wan, T. and Xie, M.},
    journal=IEEE_J_AP, 
    title={{$\mathcal{H}$-}Matrix-Based Direct Solver of {JMCFIE} for the Analysis of Scattering From Penetrable Objects}, 
    year={2022},
    month={Sep.},
    volume={70},
    number={9},
    pages={8654-8659},
    doi={10.1109/TAP.2022.3161554}
}

@article{Oijala2005,
    author={Yl\"{a}-Oijala, P. and Taskinen, M. and J\"{a}rvenp\"{a}\"{a}, S.},
    journal={Radio Sci.},
    title={Surface integral equation formulations for solving electromagnetic scattering problems with iterative methods},
    year={2005},
    volume={40},
    number={6},
    month={Nov.},
    pages={1-19},
    doi={10.1029/2004RS003169}
}

@article{Oijala2006,
    author={Taskinen, M. and Yl\"{a}-Oijala, P.},
    journal=IEEE_J_AP,
    title={Current and Charge Integral Equation Formulation},
    year={2006},
    volume={54},
    number={1},
    month={Jan.},
    pages={58-67},
    doi={10.1109/TAP.2005.861580}
}

@ARTICLE{Oijala2013,
    author={Ylä-Oijala, Pasi and Kiminki, Sami P. and Markkanen, Johannes and Järvenpää, Seppo},
    journal=IEEE_M_AP, 
    title={Error-controllable and well-conditioned {MoM} solutions in computational electromagnetics: ultimate surface integral-equation formulation}, 
    year={2013},
    month={Dec.},
    volume={55},
    number={6},
    pages={310-331},
    doi={10.1109/MAP.2013.6781749}
}

@article{Hackbusch1999,
    author = {Hackbusch, W. },
    doi = {10.1007/s006070050015},
    isbn = {1436-5057},
    journal = {Computing},
    number = {2},
    month={Apr.},
    pages = {89--108},
    title = {A Sparse Matrix Arithmetic Based on {$\mathcal{H}$-}Matrices. {Part~I}: Introduction to {$\mathcal{H}$-}Matrices},
    volume = {62},
    year = {1999},
}

@InProceedings{Hackbusch2000,
    author="Hackbusch, W.
    and Khoromskij, B.
    and Sauter, S. A.",
    title="On {H2}-Matrices",
    booktitle="Lectures on Applied Mathematics",
    year="2000",
    publisher="Springer Berlin Heidelberg",
    address="Berlin, Heidelberg",
    pages="9--29",
}

@article{Hackbusch2004,
    author = {Hackbusch, W. and Khoromskij, B. N. and Kriemann, R.},
    doi = {10.1007/s00607-004-0080-4},
    journal = {Computing},
    month = {Oct.},
    number = {3},
    pages = {207--243},
    title = {Hierarchical Matrices Based on a Weak Admissibility Criterion},
    volume = {73},
    year = {2004},
}

@article{Pals2006,
    author = {Chandrasekaran, S. and  Gu, M. and  Pals, T.},
    title = {A Fast {$ULV$} Decomposition Solver for Hierarchically Semiseparable Representations},
    journal = {SIAM J. Matrix Anal. Appl.},
    volume = {28},
    number = {3},
    pages = {603-622},
    year = {2006},
    doi = {10.1137/S0895479803436652},
}

@article{AmbikasaranDarve2013,
    author = {Ambikasaran, Sivaram and Darve, Eric},
    doi = {10.1007/s10915-013-9714-z},
    journal = {J. Sci. Comput.},
    month = {Dec.},
    number = {3},
    pages = {477--501},
    title = {An {$\mathcal{O}(N \log N)$} Fast Direct Solver for Partial Hierarchically Semi-Separable Matrices},
    volume = {57},
    year = {2013},
}

@article{Greengard2016,
    author = {Vico, Felipe and Ferrando, Miguel and Greengard, Leslie and Gimbutas, Zydrunas},
    title = {The Decoupled Potential Integral Equation for Time-Harmonic Electromagnetic Scattering},
    journal = {Commun. Pure Appl. Math.},
    volume = {69},
    number = {4},
    month={Apr.},
    pages = {771-812},
    doi = {https://doi.org/10.1002/cpa.21585},
    year = {2016}
}

@article{Greengard2018,
    author = {Felipe Vico and Leslie Greengard and Miguel Ferrando},
    title = {Decoupled field integral equations for electromagnetic scattering from homogeneous penetrable obstacles},
    journal = {Commun. Partial Differ. Equ.},
    volume = {43},
    number = {2},
    pages = {159-184},
    month={Mar.},
    year = {2018},
    publisher = {Taylor \& Francis},
    doi = {10.1080/03605302.2018.1446447},
}

@Article{Chew2014,
    author={Weng Cho Chew},
    title={Vector Potential Electromagnetics with Generalized Gauge for Inhomogeneous Media: Formulation},
    volume={149},
    journal={Prog. Electromagn. Res.},
    year={2014},
    month={Sep.},
    pages={69-84},
    doi={10.2528/PIER14060904}
}

@article{Chew2000,
    author={Jun-Sheng Zhao and Weng Cho Chew},
    journal=IEEE_J_AP,
    title={Integral equation solution of {M}axwell's equations from zero frequency to microwave frequencies},
    year={2000},
    volume={48},
    number={10},
    month={Oct.},
    pages={1635-1645},
    doi={10.1109/8.899680}
}

@article{Chew2008,
    author = {Qian, Zhi Guo and Chew, Weng Cho},
    title = {A quantitative study on the low frequency breakdown of {EFIE}},
    journal = {Microw. Opt. Technol. Lett.},
    volume = {50},
    number = {5},
    month={Mar.},
    pages = {1159-1162},
    doi = {https://doi.org/10.1002/mop.23324},
    year = {2008}
}

@article{Chew2007,
    author={Qian, Zhi Guo and Chew, Weng Cho and Suaya, Roberto},
    journal=IEEE_J_MTT,
    title={Generalized Impedance Boundary Condition for Conductor Modeling in Surface Integral Equation},
    year={2007},
    volume={55},
    number={11},
    month={Nov.},
    pages={2354-2364},
    doi={10.1109/TMTT.2007.908678}
}

@article{Chew2009,
    author={Qian, Zhi-Guo and Chew, Weng Cho},
    journal=IEEE_J_AP,
    title={Fast Full-Wave Surface Integral Equation Solver for Multiscale Structure Modeling},
    year={2009},
    volume={57},
    number={11},
    month={Nov.},
    pages={3594-3601},
    doi={10.1109/TAP.2009.2023629}
}

@article{QianChew2009,
    author={Zhiguo Qian and Mei Song Tong and Weng Cho Chew},
    title={Conductive Medium Modeling with an Augmented {GIBC} Formulation},
    volume={99},
    journal={Prog. Electromagn. Res.},
    year={2009},
    month={Nov.},
    pages={261-272},
    doi={10.2528/PIER09100702}
}

@ARTICLE{Chew2017,
    author={Xia, Tian and Gan, Hui and Wei, Michael and Chew, Weng Cho and Braunisch, Henning and Qian, Zhiguo and Aygün, Kemal and Aydiner, Alaeddin},
    journal=IEEE_J_AP,
    title={An Integral Equation Modeling of Lossy Conductors With the Enhanced Augmented Electric Field Integral Equation},
    year={2017},
    volume={65},
    number={8},
    month={Aug.},
    pages={4181-4190},
    doi={10.1109/TAP.2017.2718587}
}

@article{Andriulli2008,
    author={Andriulli, Francesco P. and Cools, Kristof and Bagci, Hakan and Olyslager, Femke and Buffa, Annalisa and Christiansen, Snorre and Michielssen, Eric},
    journal=IEEE_J_AP,
    title={A Multiplicative {C}alderon Preconditioner for the Electric Field Integral Equation},
    year={2008},
    volume={56},
    number={8},
    month={Aug.},
    pages={2398-2412},
    doi={10.1109/TAP.2008.926788}
}

@article{Andriulli2011,
    author={Cools, Kristof and Andriulli, Francesco P. and Michielssen, Eric},
    journal=IEEE_J_AP,
    title={A {C}alderón Multiplicative Preconditioner for the {PMCHWT} Integral Equation},
    year={2011},
    volume={59},
    number={12},
    month={Dec.},
    pages={4579-4587},
    doi={10.1109/TAP.2011.2165465}
}

@article{Andriulli2013,
    author={Andriulli, Francesco P. and Cools, Kristof and Bogaert, Ignace and Michielssen, Eric},
    journal=IEEE_J_AP,
    title={On a Well-Conditioned Electric Field Integral Operator for Multiply Connected Geometries},
    year={2013},
    volume={61},
    month={Apr.},
    number={4},
    pages={2077-2087},
    doi={10.1109/TAP.2012.2234072}
}

@article{Andriulli2017,
    author={Beghein, Yves and Mitharwal, Rajendra and Cools, Kristof and Andriulli, Francesco P.},
    journal=IEEE_J_AP,
    title={On a Low-Frequency and Refinement Stable {PMCHWT} Integral Equation Leveraging the Quasi-{H}elmholtz Projectors},
    year={2017},
    month={Oct.},
    volume={65},
    number={10},
    pages={5365-5375},
    doi={10.1109/TAP.2017.2738061}
}

@article{Andriulli2021,
    author={Adrian, Simon B. and D\'ely, Alexandre and Consoli, Davide and Merlini, Adrien and Andriulli, Francesco P.},
    journal={IEEE Open J. Antennas Propag.},
    title={Electromagnetic Integral Equations: Insights in Conditioning and Preconditioning},
    year={2021},
    volume={2},
    number={},
    month={Oct.},
    pages={1143-1174},
    doi={10.1109/OJAP.2021.3121097}
}

@ARTICLE{Andriulli2023,
    author={Merlini, Adrien and Henry, Clément and Consoli, Davide and Rahmouni, Lyes and Dély, Alexandre and Andriulli, Francesco P.},
    journal=IEEE_J_AP,
    title={Laplacian Filtered Loop-Star Decompositions and Quasi-{H}elmholtz Filters: Definitions, Analysis, and Efficient Algorithms},
    year={2023},
    volume={71},
    number={12},
    month={Dec.},
    pages={9289-9302},
    doi={10.1109/TAP.2023.3283043}
}

@article{Andriulli2012,
    author={Andriulli, Francesco P.},
    journal=IEEE_J_AP,
    title={Loop-Star and Loop-Tree Decompositions: Analysis and Efficient Algorithms},
    year={2012},
    volume={60},
    month={May},
    number={5},
    pages={2347-2356},
    doi={10.1109/TAP.2012.2189723}
}

@INPROCEEDINGS{AndriulliConf2018,
    author={Chhim, Tiffany L. and Adrian, Simon B. and Andriulli, Francesco P.},
    booktitle={Proc. IEEE Int. Symp. Antennas Propag. (APS/URSI)}, 
    title={On the Spectral Behavior and Normalization of a Resonance-Free and High-Frequency Stable Integral Equation}, 
    year={2018},
    volume={},
    month={Jul.},
    number={},
    pages={2447-2448},
    doi={10.1109/APUSNCURSINRSM.2018.8608769}
}

@INPROCEEDINGS{AndriulliConf2019,
    author={Dély, Alexandre and Merlini, Adrien and Adrian, Simon B. and Andriulli, Francesco P.},
    booktitle={Proc. Int. Conf. Electromagn. Adv. Appl. (ICEAA)}, 
    title={On Preconditioning Electromagnetic Integral Equations in the High Frequency Regime via {Helmholtz} Operators and quasi-{Helmholtz} Projectors}, 
    year={2019},
    volume={},
    number={},
    month={Sep.},
    pages={1338-1341},
    doi={10.1109/ICEAA.2019.8879366}
}

@article{Turc2014,
    title = {Well-conditioned boundary integral equation formulations for the solution of high-frequency electromagnetic scattering problems},
    journal = {Comput. Math. Appl.},
    volume = {67},
    number = {10},
    pages = {1772-1805},
    year = {2014},
    month = {Jun.},
    doi = {https://doi.org/10.1016/j.camwa.2014.04.003},
    author = {Yassine Boubendir and Catalin Turc},
}

@article{Volkov2014,
    title = {An All-Frequency Weakly-Singular surface integral equation for electromagnetism in dielectric media: Reformulation and well-posedness analysis},
    journal = {J. Math. Anal. Appl.},
    volume = {412},
    number = {1},
    pages = {277-300},
    month={Apr.},
    year = {2014},
    issn = {0022-247X},
    doi = {https://doi.org/10.1016/j.jmaa.2013.10.059},
    author = {M. Ganesh and S. C. Hawkins and D. Volkov},
}

@article{Shanker2022,
    author={Baumann, Luke and Aktulga, H. M. and Macon, Charles A. and Shanker, B.},
    journal=IEEE_J_AP,
    title={Decoupled Potential Integral Equation for Electromagnetic Scattering From Arbitrarily Shaped Dielectric Objects},
    year={2022},
    volume={70},
    month={Aug.},
    number={8},
    pages={6858-6871},
    doi={10.1109/TAP.2022.3161278}
}

@article{Shanker2019,
    author={Li, Jie and Fu, Xin and Shanker, Balasubramaniam},
    journal=IEEE_J_AP,
    title={Decoupled Potential Integral Equations for Electromagnetic Scattering From Dielectric Objects},
    year={2019},
    volume={67},
    month={Mar.},
    number={3},
    pages={1729-1739},
    doi={10.1109/TAP.2018.2883636}
}

@article{Zutter2005,
    author={De Zutter, D. and Knockaert, L.},
    journal=IEEE_J_MTT,
    title={Skin effect modeling based on a differential surface admittance operator},
    year={2005},
    volume={53},
    number={8},
    month={Aug.},
    pages={2526-2538},
    doi={10.1109/TMTT.2005.852766}
}

@article{Zutter2025,
    author={Pattyn, Tim and Sun, Xiao and Beyne, Eric and De Zutter, Daniël and Huynen, Martijn and Vande Ginste, Dries},
    journal=IEEE_J_MTT,
    title={Broadband Electromagnetic Modeling of On-Chip Passives Using a Differential Surface Admittance Operator for {3-D} Piecewise Homogeneous Structures},
    year={2025},
    volume={73},
    number={12},
    pages={10199-10212},
    month={Dec.},
    doi={10.1109/TMTT.2025.3600958}
}

@article{Utkarsh2016,
    author={Patel, Utkarsh R. and Triverio, Piero},
    journal=IEEE_J_MTT,
    title={Skin Effect Modeling in Conductors of Arbitrary Shape Through a Surface Admittance Operator and the Contour Integral Method},
    year={2016},
    volume={64},
    number={9},
    month={Sep.},
    pages={2708-2717},
    doi={10.1109/TMTT.2016.2593721}
}

@ARTICLE{ZhaoJFLee2005,
    author={Kezhong Zhao and Vouvakis, M. N. and Jin-Fa Lee},
    journal=IEEE_J_EMC, 
    title={The adaptive cross approximation algorithm for accelerated method of moments computations of {EMC} problems}, 
    month={Nov.},
    year={2005},
    volume={47},
    number={4},
    pages={763-773},
    doi={10.1109/TEMC.2005.857898}
}

@ARTICLE{Mao2018,
    author={Zhao, Yu and Mao, Junfa},
    journal=IEEE_J_MTT, 
    title={Equivalent Surface Impedance-Based Mixed Potential Integral Equation Accelerated by Optimized $\cal {H}$-Matrix for {3-D} Interconnects}, 
    year={2018},
    volume={66},
    number={1},
    month={Jan.},
    pages={22-34},
    doi={10.1109/TMTT.2017.2731956}
}

@article{DJiao2013,
    author={Chai, Wenwen and Jiao, Dan},
    journal={Proc. IEEE},
    title={Direct Matrix Solution of Linear Complexity for Surface Integral-Equation-Based Impedance Extraction of Complicated {3-D} Structures},
    year={2013},
    volume={101},
    month={Feb.},
    number={2},
    pages={372-388},
    doi={10.1109/JPROC.2012.2190577}
}

@ARTICLE{DJiao2014,
    author={Zhu, Jianfang and Omar, Saad and Jiao, Dan},
    journal=IEEE_J_AP,
    title={Solution of the Electric Field Integral Equation When It Breaks Down},
    year={2014},
    volume={62},
    month={Aug.},
    number={8},
    pages={4122-4134},
    doi={10.1109/TAP.2014.2322899}
}

@ARTICLE{Roth2023,
    author={Roth, Thomas E. and Ma, Ruichao and Chew, Weng C.},
    journal=IEEE_M_AP,
    title={The Transmon Qubit for Electromagnetics Engineers: An introduction},
    year={2023},
    volume={65},
    month={Apr.},
    number={2},
    pages={8-20},
    doi={10.1109/MAP.2022.3176593}
}

@article{Oliver2019,
    author = {Krantz, P. and Kjaergaard, M. and Yan, F. and Orlando, T. P. and Gustavsson, S. and Oliver, W. D.},
    title = {A quantum engineer's guide to superconducting qubits},
    journal = {Appl. Phys. Rev.},
    volume = {6},
    number = {2},
    month = {Jun.},
    pages = {021318},
    year = {2019},
    issn = {1931-9401},
    doi = {10.1063/1.5089550},
}

@article{gmsh,
    author = {Geuzaine, Christophe and Remacle, Jean-François},
    title = {Gmsh: A {3-D} finite element mesh generator with built-in pre- and post-processing facilities},
    journal = {Int. J. Numer. Methods Eng.},
    volume = {79},
    number = {11},
    month = {May},
    pages = {1309-1331},
    doi = {https://doi.org/10.1002/nme.2579},
    year = {2009}
}

@ARTICLE{skrf_mini,
    author={{A. Arsenovic \emph{et al.}}},
    journal=IEEE_M_MW,
    title={{scikit-rf}: An Open Source {P}ython Package for Microwave Network Creation, Analysis, and Calibration},
    year={2022},
    volume={23},
    number={1},
    month={Jan.},
    pages={98-105},
    doi={10.1109/MMM.2021.3117139}
}

@misc{hfss,
    title = {Ansys {HFSS}},
    howpublished = {\url{https://www.ansys.com/en-gb/products/electronics/ansys-hfss}},
    note = {Accessed: Jan. 27, 2026}
}

@misc{maxwell,
    title = {Ansys {Maxwell}},
    howpublished = {\url{https://ansys.synopsys.com/en-gb/products/electronics/ansys-maxwell}},
    note = {Accessed: Nov. 15, 2026}
}

@misc{OpenBEM,
    author = {Shashwat Sharma},
    title = {{OpenBEM} (\url{github.com/shash-sharma/OpenBEM})},
    howpublished = {Mar. 2026 [Online]},
    note = {{Available}: \url{https://doi.org/10.5281/zenodo.19337991}}
}

@article{JMMCTFSIE,
    author={Shashwat Sharma},
    title = {A Combined-Field Boundary Element Formulation for the Electromagnetic Modeling of Multiscale Interconnects and Integrated Passive Components from {DC} to Millimeter-Wave Frequencies},
    journal = IEEE_J_MMCT,
    volume={11},
    month={Jul.},
    number={},
    pages={437-452},
    year = {2026},
    doi = {10.1109/JMMCT.2026.3715919}
}

@article{AWPLSLIM,
    author={Shashwat Sharma and Piero Triverio},
    journal=IEEE_J_AWPL,
    title={{SLIM}: A Well-Conditioned Single-Source Boundary Element Method for Modeling Lossy Conductors in Layered Media},
    year={2020},
    volume={19},
    number={12},
    pages={2072-2076},
    month={Dec.},
    doi={10.1109/LAWP.2020.3022593}
}

@article{MTTVPIE,
    author={Sharma, Shashwat and Triverio, Piero},
    journal=IEEE_J_MTT,
    title={Electromagnetic Modeling of Lossy Interconnects From {DC} to High Frequencies With a Potential-Based Boundary Element Formulation},
    year={2022},
    volume={70},
    number={8},
    month={Aug.},
    pages={3847-3861},
    doi={10.1109/TMTT.2022.3180390}
}

@article{AWPLVPIE,
    author={Sharma, Shashwat and Triverio, Piero},
    journal=IEEE_J_AWPL,
    title={Electromagnetic Modeling of Lossy Materials With a Potential-Based Boundary Element Method},
    year={2022},
    volume={21},
    month={Feb.},
    number={2},
    pages={391-395},
    doi={10.1109/LAWP.2021.3132626}
}

@article{JMMCTSLDM,
    author={Sharma, Shashwat and Triverio, Piero},
    journal=IEEE_J_MMCT,
    title={A Single-Layer Dual-Mesh Boundary Element Method for Multiscale Electromagnetic Modeling of Penetrable Objects in Layered Media},
    year={2021},
    volume={6},
    number={},
    month={Oct.},
    pages={158-170},
    doi={10.1109/JMMCT.2021.3122427}
}

@ARTICLE{TAPAIMin,
    author={Shashwat Sharma and Piero Triverio},
    journal=IEEE_J_AP,
    title={An Accelerated Surface Integral Equation Method for the Electromagnetic Modeling of Dielectric and Lossy Objects of Arbitrary Conductivity},
    year={2021},
    volume={69},
    number={9},
    pages={5822-5836},
    month={Sep.},
    doi={10.1109/TAP.2021.3061119}
}

\vfill

\end{document}